# Large language models for post-publication research evaluation: Evidence from expert recommendations and citation indicators

Mengjia Wu[1], Yi Zhang[1], Robin Haunschild[2], and Lutz Bornmann[2,3]

[1] *Mengjia.Wu@student.uts.edu.au, Yi.Zhang@uts.edu.au*
Australian Artificial Intelligence Institute, Faculty of Engineering and Information Technology, University of Technology Sydney (Australia)

[2] *R.Haunschild@fkf.mpg.de, L.Bornmann@fkf.mpg.de*
Max Planck Institute for Solid State Research, Heisenbergstr. 1, 70569 Stuttgart (Germany)

[3] *bornmann@gv.mpg.de*
Science Policy and Strategy Department, Administrative Headquarters of the Max Planck Society, Hofgartenstr. 8, 80539 Munich (Germany)

**Abstract**

Assessing the quality of scientific research is essential for scholarly communication, yet widely used approaches face limitations in scalability, subjectivity, and time delay. Recent advances in large language models (LLMs) offer new opportunities for automated research evaluation based on textual content. This study examines whether LLMs can support post-publication peer review tasks by benchmarking their outputs against expert judgments and citation-based indicators. Two evaluation tasks are constructed using articles from the H1 Connect platform: identifying high-quality articles and performing finer-grained evaluation including article rating, merit classification, and expert style commenting. Multiple model families, including BERT models, general-purpose LLMs, and reasoning oriented LLMs, are evaluated under multiple learning strategies. Results show that LLMs perform well in coarse grained evaluation tasks, achieving accuracy above 0.8 in identifying highly recommended articles. However, performance decreases substantially in fine-grained rating tasks. Few-shot prompting improves performance over zero-shot settings, while supervised fine-tuning produces the strongest and most balanced results. Retrieval augmented prompting improves classification accuracy in some cases but does not consistently strengthen alignment with citation indicators. The overall correlations between model outputs and citation indicators remain positive but moderate.

# 1. Introduction

Peer review plays a critical role in shaping scholarly communication, resource allocation, and knowledge diffusion (Bornmann, 2008). Beyond traditional peer review, expert-driven recommendation platforms and citation-based indicators have become widely used to assess the quality and impact of scientific articles. However, both expert judgment and citation metrics face well-documented limitations: expert evaluations are costly, subjective, and difficult to scale, while citations are delayed, field-dependent, and only imperfectly reflect intellectual merit. These challenges have motivated increasing interest in computational approaches to research evaluation within information science.

Recent advances in large language models (LLMs) have introduced new possibilities for automated research evaluation. Exemplified by OpenAI's GPT series and Google's BERT, LLMs have demonstrated remarkable capabilities in natural language understanding and generation (Farhat et al., 2023). Trained with large-scale textual corpora and capable of contextual reasoning over scientific language, LLMs offer the potential to assess scholarly content directly from article titles and abstracts, rather than relying on ex post signals such as citation counts. Early studies suggest that LLMs can support tasks such as research quality identification (Thelwall, 2024a), peer review assistance (Liang et al., 2024), manuscript screening (López-Pineda et al., 2025), and general complex research evaluation tasks (Thelwall, 2025). Yet, fundamental questions remain unresolved regarding how well LLM-based evaluations compare with human experts, how LLMs' results relate to established citation indicators, and what forms of learning or supervision strategies are required for reliable performance.

In this study, we address three interrelated concerns that remain insufficiently explored in prior research. Specifically, we examine (1) the extent to which both general-purpose and reasoning LLMs can support post-publication research evaluation, (2) the conditions and learning strategies under which LLMs can achieve reliable and acceptable performance, and (3) how LLM-based evaluations compare with established citation-based indicators. Accordingly, this study seeks to answer the following research questions:

- **Research question 1 (RQ1):** To what extent can LLMs perform post-publication research evaluation tasks in comparison with expert judgments?
- **Research question 2 (RQ2):** Which LLM and learning strategies yield the most reliable performance for post-publication research evaluation?

- **Research question 3 (RQ3):** When benchmarked against citation-based indicators, how do model selection and learning strategies influence performance?

To address the three proposed research questions, we perform post-publication research evaluation through two complementary tasks: Task 1, high-quality article identification, and Task 2, article rating, merit classification, and expert-style commenting. Task 1 focuses on distinguishing high- and low-quality articles from a mixed pool, whereas Task 2 targets more fine-grained evaluative behaviors, including rating articles, assigning merit classification codes, and generating qualitative comments akin to those produced by human experts. Both tasks are grounded in real-world evaluation practices of biomedical faculty researchers from H1 Connect (formerly Faculty Opinions).

To examine the impact of model architecture, we evaluate three representative tracks of language models: encoder-based models (BERT and its variants), general-purpose LLMs, and reasoning-oriented LLMs. Together, these model families cover the primary types of language models currently used by both the public and scientific research communities. For both Task 1 and Task 2, all models are evaluated under in-context learning and supervised fine-tuning settings. Model outputs are compared against expert annotations and citation-based indicators to assess the extent to which LLM-based evaluations align with human judgment and citation performance.

The experimental results yield several key findings. First, LLMs demonstrate strong performance in coarse-grained evaluation tasks, achieving accuracy above 0.8 in differentiating high- and low-quality articles in Task 1. However, they struggle with finer-grained stratification of article quality when only abstract-level semantic information is available, as reflected by rating accuracy below 0.5 in Task 2. Second, for more complex evaluation tasks such as merit classification and expert-style commenting in Task 2, models from the Gemini and Llama families achieve competitive performance and emerge as viable alternatives to the GPT series. Third, both few-shot learning and supervised fine-tuning effectively improve alignment between LLM outputs and human annotations. While semantically similar examples can enhance recommendation and rating accuracy, they do not consistently strengthen alignment with citation-based indicators, suggesting that semantic similarity alone may be insufficient to capture citation-related signals. Finally, reasoning-oriented LLMs do not, in general, exhibit clear advantages over general-purpose LLMs across the evaluated tasks. An exception is GPT-5, which demonstrates consistently strong performance in both tasks; however, due to its closed-source nature, further analysis of the underlying causes of this advantage is not possible.

The rest of this paper is organized as follows: The relevant work of research evaluation with LLMs is reviewed and summarized in Section 2. Section 3 details our methodological framework, model and learning strategy selection, and experimental settings. Section 4 profiles the experimental results across dimensions of human opinion alignment and citation correlation analysis, further detailing the results of data efficiency analysis. The last section, Section 5, wraps up this paper with conclusions and discussion.

## 2. Related work

In recent years, the development of language models has substantially advanced text representation and classification in scholarly domains. Initially, encoder-only models such as BERT, SciBERT, BioBERT, and PubMedBERT have demonstrated strong performance in scientific text classification and information extraction tasks (Wu et al., 2025a). Their effectiveness is often attributed to domain-adaptive pretraining and supervised fine-tuning on task-specific data. These models have been widely applied in computational research evaluation settings, where textual signals from abstracts or full texts are used to approximate expert judgments of relevance or quality.

More recently, the rise of generative LLMs, exemplified by ChatGPT (Farhat et al., 2023), has been explored for a broader range of scholarly assessment tasks within the interest of the information science community, including scholarly recommendation (Chen et al., 2026; Jia et al., 2025), research impact prediction (Lu et al., 2025), societal impact assessment (Kousha & Thelwall, 2025; Wu et al., 2025a), and novelty evaluation (Wu et al., 2025c). Several empirical studies have examined the feasibility of using LLMs in peer review (Liang et al., 2024; Liu & Shah, 2023; López-Pineda et al., 2025; Zhu et al., 2026). Collectively, this stream of work has provided evidence that LLMs can provide useful support for well-defined reviewing tasks, while still facing limitations in inaccurate and subjective evaluations.

More specifically, early investigations focused on task-specific performance. For example, Liu and Shah (2023) evaluated GPT-4 on structured reviewing tasks such as identifying factual errors, verifying checklist items, and selecting the stronger paper from pairs of abstracts. Their results show that LLMs perform well on objective and rule-based tasks but struggle when judgments require nuanced assessments of research quality. This finding indicates that LLMs are better suited as assistants for constrained evaluative subtasks rather than as substitutes for human reviewers.

Subsequent work has examined the quality and usefulness of LLM-generated feedback on a scale. Liang et al. (2024) conducted a large empirical analysis comparing GPT-4-generated feedback with human peer review reports across thousands of submissions from leading journals and conferences. They observed substantial overlap between the issues raised by the model and those identified by human reviewers, particularly for lower-quality papers. A complementary user study further revealed that researchers generally found LLM-generated feedback helpful, suggesting that such models may be especially valuable in under-resourced research environments.

Subsequent studies have extended this line of inquiry toward quality estimation. Thelwall and Yaghi (2024) assessed whether ChatGPT-4o-mini could approximate expert quality judgments by comparing its scores, based solely on titles and abstracts, with departmental averages from the United Kingdom's Research Excellence Framework 2021. Their findings show generally positive correlations, with stronger alignment in the physical and health sciences. However, these results also highlight disciplinary variability and the constraints of abstract-level assessment, underscoring the need for cautious interpretation.

Despite these promising results, existing studies also consistently emphasize the limitations of LLM-based evaluation (Sun, 2025). LLMs are known to generate plausible but inaccurate content, raising concerns about hallucinated information in evaluative contexts (Thelwall, 2024b). As a result, current evidence supports the view that LLMs can meaningfully augment peer review workflows but are not yet reliable as standalone evaluators. In a more concrete context, it is worth investigating how current LLMs perform when compared against human and citation metrics in post-publication tasks, as well as in what conditions the LLMs can achieve acceptable results.

## 3. Data and Methodology

The overall research framework is illustrated in Figure 1. Three tracks of LLMs are selected and tested in this study, with zero-shot, three few-shot strategies, and supervised fine-tuning learning strategies employed and applied. The data collection and curation process are described in Section 3.1. Descriptions of task details, selected models, and model learning strategies are detailed in Sections 3.2-3.4.

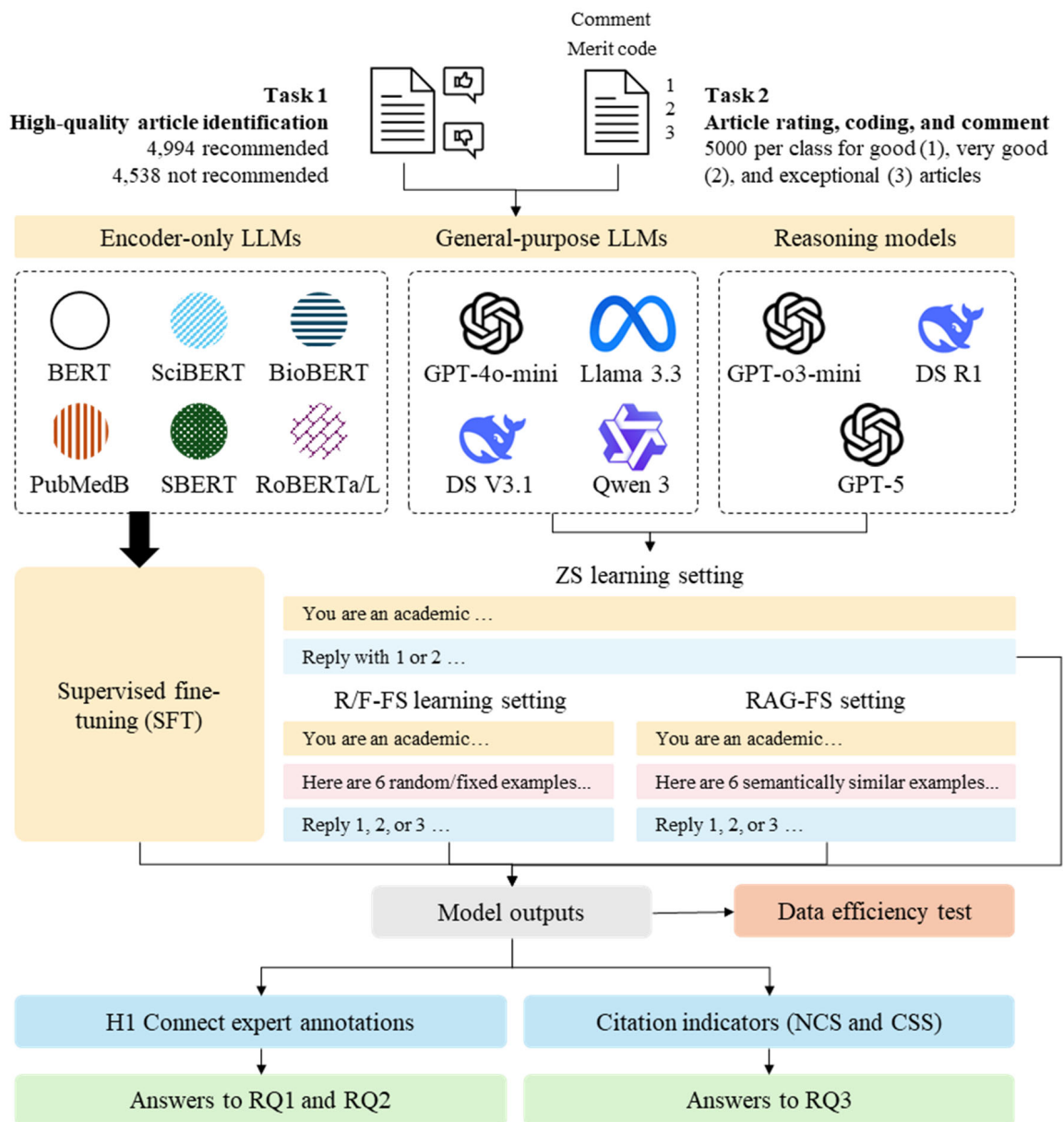

**Figure 1. The methodological framework**

*3.1 Task formulation and data collection*

To answer the proposed RQs in practical post-publication evaluation scenarios, we curated two representative evaluation tasks: High-quality article identification (Task 1) and recommended article rating, coding, and commenting (Task 2). The former task focuses on differentiating high-quality articles, which are defined as recommended by at least three experts from H1 Connect, from those that receive no recommendations. The latter task centers on more nuanced differences of those recommended articles, with the aim of rating those articles accurately (which is defined by the number of experts recommending this article) and providing justifications – why it is worth recommending. Specifically, the task settings and data collection details are given below.

**Task 1: High-quality article identification**

This task aims to evaluate how effectively LLMs can identify high-quality articles from a mixed pool of articles, compared to the judgment of human experts. Low-quality articles are defined as those with no expert recommendations and high-quality articles are those with three or more expert recommendations. To construct a mixed pool for testing, we compiled 4,538 articles from OpenAlex (Priem et al., 2022) – a bibliographic catalogue of scientific papers – with no expert recommendations and 4,994 articles with three or more expert recommendations. The not-recommended articles were published between 2010 and 2020 in the same journal, with the same volume and issue as the recommended articles. We excluded the journals *Science*, *Nature*, *Proceedings of the National Academy of Sciences of the United States of America*, *Science Advances*, *Nature Communications*, *Scientific Reports*, and *PLOS ONE* due to their multidisciplinary nature for the selection of not-recommended articles. The selected LLMs are required to retrieve the 4,994 high-quality articles from this pool as accurately as possible.

**Task 2: Recommended article rating, coding, and commenting**

In addition to Task 1, a more curated and sophisticated task is designed to evaluate if LLMs can identify more nuanced quality differences (ratings, merit codes, and comments) of the articles recommended by human experts. To avoid complications in synthesizing expert ratings, we focused on articles with only one recommendation at this stage, but the recommendation has been marked by human experts as good (rating 1), very good (rating 2), or excellent (rating 3), indicating three rating levels. In addition to the ratings, each article also comes with one or more merit codes and commentary annotated by experts. The data collection also follows procedures as in Task 1, resulting in 86,805 articles with a rating of 1, 54,154 articles with a rating of 2, and 11,089 articles with a rating of 3 (roughly 8:5:1). Considering the computational costs for model testing, we sampled a balanced dataset that consists of 5,000 articles from each rating of 1, 2, and 3, ending up with 15,000 articles. In this setting, the titles and abstracts of articles are fed into LLMs to output the rating. For the LLMs with generative capabilities, they can also provide specific merit codes, and commentaries. The merit code distribution across ratings (1, 2, and 3) for the 15,000 articles is shown in Figure 2.

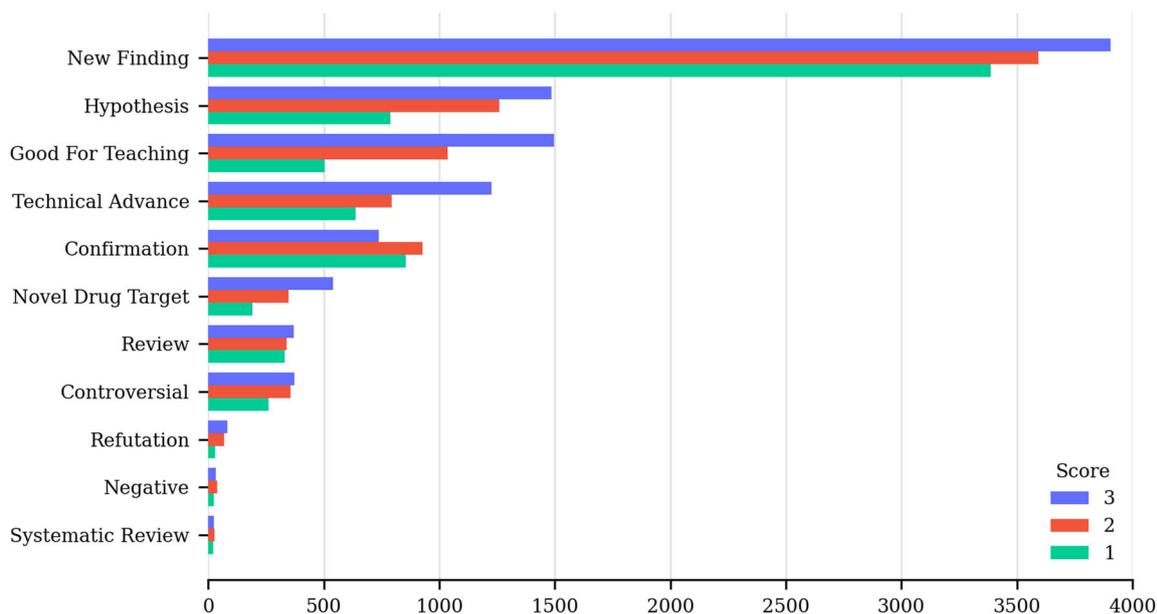

**Figure 2. Distribution of article merit codes across the three rating levels**

*3.2 Selected models*

Three tracks of LLMs are selected and evaluated in this study. Their architecture, along with the justifications, are detailed as follows:

- BERT-based models: BERT-based models are a series of representative encoder-only language models that are specified for text classification and understanding tasks. The effectiveness of BERT-based models in text classification and understanding has been recognized in multiple pilot studies (Wu et al., 2025a). Models in this track used in this study include BERT (Devlin et al., 2019), SciBERT (Beltagy et al., 2019), BioBERT (Lee et al., 2020), SentenceBERT (SBERT) (Reimers & Gurevych, 2019), RoBERTa (Liu et al., 2019), RoBERTa-large, and PubMedBERT (Gu et al., 2021). Within this track, both Task 1 and Task 2 rating are performed as classification tasks, while merit code and comment generation in Task 2 are skipped off for BERT-based models.

- General-purpose LLMs: This track of models is a general-purpose tool for open-ended research evaluation. In this track, we selected several representative LLM families developed by worldwide leading big IT heads: Qwen from Alibaba (Yang et al., 2025), Llama models (Touvron et al., 2023) from Meta (previously Facebook), and GPT models (Achiam et al., 2023) from OpenAI. All the LLMs can generate corresponding answers according to natural language prompts provided by users.

- Reasoning LLMs: This specialized category of LLMs, known as reasoning models, is designed to tackle complex, multi-step reasoning tasks such as solving advanced mathematical problems and coding tasks with reasoning modules and self-verification embedded in their architecture. To examine the performance of reasoning models on post-publication peer review tasks, multiple representative reasoning LLMs, GPT-o3-mini, and DeepSeek R1, and GPT-5, are also included in the analyses.

*3.3 Model learning strategies*

Currently, there are multiple strategies to guide LLMs to optimize LLM outputs and adapt to specific tasks. For each learning strategy, both datasets were randomly divided into training sets and test sets according to a ratio of 8:2. All the articles with their annotations (recommendation, rating, merit code, or commentary) exposed to LLMs are from the training sets (7,625 articles in dataset 1 and 12,000 articles in dataset 2), then the reported results are all based on the test sets, with 1,907 articles from dataset 1 (1,016 recommended and 891 not recommended) and 3,000 articles (1,037 with rating 1, 970 with rating 2, and 993 with rating 3) from dataset 2. In this study, we selected and utilized the following mainstream LLM learning strategies:

*3.3.1. In context learning (ICL)*

In-context learning (ICL) (Dong et al., 2024) is a prompt-engineering technique designed for both general-purpose and reasoning LLMs. It works by providing contextual information, sometimes along with task-specific input-output pair demonstrations directly in the prompts, enabling models to generate responses for given questions. ICL does not alter the model's parameters but just modifies the prompts to achieve more accurate outputs. This makes ICL a low-cost and user-friendly approach to leveraging LLMs. In this study, we selected the following representative ICL approaches:

- Zero-shot (ZS) learning: ZS learning is a prompt engineering scheme that only provides the descriptions of the research evaluation task itself. This strategy does not change the model configuration or parameters but just instructs the models to provide output based on the prompt given.
- Few-shot (FS) learning: FS learning aims to guide LLMs to learn patterns of resolving the task with a few labelled examples. In addition to the prompt provided in ZS learning, this strategy feeds examples to LLMs, that is in our study, article abstracts with human expert annotations, to teach LLMs how to perform research evaluation like a real human expert. Our pilot research (Wu et al., 2025b) has proven that FS is a possible strategy for enhancing the output quality. In this study, we continue to provide six labeled examples for each target sample. According to the ways that examples were selected, we created the following different strategies:

- Fixed-FS (F-FS): For each target article, F-FS provides the same fixed six example articles (three recommended and three not recommended examples for Task 1, two of each articles with rating 1, 2, and 3 for Task 2) from the training set in addition to the instruction prompt.
- Random-FS (R-FS): For each target article, the six example articles with the same label distribution with F-FS were randomly selected from the training set to avoid potential bias of the selected examples.
- Retrieval-augmented generation - FS (RAG-FS) (Wang et al., 2025): For each article, this strategy selects top six semantically similar articles to the target article. This strategy was shaped based on the assumption that semantically similar articles should demonstrate more useful patterns for the LLMs to learn to correctly classify the target article. The semantic similarity was measured by PubMedBERT, as it has been proven to be one of the most useful universal text representation models in the biomedical field (Gu et al., 2021).

### 3.3.2. Supervised fine-tuning (SFT)

SFT adapts LLMs to specific tasks by retraining model parameters on labelled data, which in this study consists of article recommendations and ratings, while largely retaining knowledge acquired during pre-training. In contrast to ICL, SFT explicitly alters model parameters to capture task-specific data patterns, but this comes at a substantially higher computational and time cost, particularly for generative LLMs, and often results in more task-specific models with potentially reduced generalizability. Although SFT can in principle be applied to all LLMs, the associated training resource requirements limited its application in this study to BERT models and the GPT-4o-mini model, with the latter implemented via the OpenAI API[1].

### 3.4 Experimental settings and evaluation

For evaluating LLMs' outputs for Task 1 and Task 2, we further design multiple types of experiments and are designed as follows:

---

[1] https://platform.openai.com/docs/api-reference/fine-tuning

- ICL and SFT settings: For all the generative LLMs, we select model temperature as zero to guarantee the most reproducibility and less output variability. For each SFT setting, we selected training. For model training and inference, the BERT-based models were trained and tested on a high-computation performance center node with 2 x NVIDIA L4 (24GB) GPU, GPT models' inference are performed through the OpenAI API[2], Gemini models are accessed through Google's model API[3], all other LLMs are accessed via the Deep Infra platform[4].

- Human expert alignment: This reveals how LLMs can produce aligned recommendations with human experts. For both Task 1 and Task 2, the performance of giving recommendations and ratings is measured by precision, recall, and F1-score. For the merit code generated in Task 2, the average Jaccard Coefficient (Jaccard, 1912) and F1-scores are employed. For the comments in Task 2, we adopted Bilingual Evaluation Understudy (BLEU) score (Papineni et al., 2002), ROUGE-1, 2, L (Lin, 2004) scores to measure the text similarity between comments given by LLMs and human experts.

- Citation correlation: This evaluation explores how the LLMs outputs and expert opinions agree with citation indicators. Here, we collected the two citation metrics, normalized citation scores (NCS) (Scheidsteger et al., 2025) and characteristic scores and scales (CSS) (Glänzel, 2007) based on the NCS values, of all the articles in Task 1 and Task 2. Rank-biserial correlation is used for Task 1 (binary outcomes), while Spearman correlation is used for Task 2 (ordinal ratings).

- Data efficiency assessment: In the SFT strategy, partial data is required to retrain the LLMs to adapt to the research evaluation task. However, how much data can achieve satisfying testing accuracy remains a critical issue to investigate, which can provide more insights into the practicability of model utilization in real-world scenarios. Hence, we gradually reduced the training set size to 60, 40, and 20 percent of the entire dataset size to observe how the testing results change.

---

[2] https://openai.com/api/

[3] https://ai.google.dev/gemini-api/docs

[4] https://deepinfra.com/

## 4. Results

*4.1. Task 1 - High-quality article identification results*

*4.1.1. Human expert alignment*

Table 1 presents the accuracy of all evaluated LLMs under four ICL strategies and the SFT setting for Task 1. Overall, SFT consistently and substantially outperforms all ICL-based strategies across model architectures. Among the fine-tuned models, GPT-4o-mini achieves the highest accuracy (0.840), surpassing all BERT-based models and improving upon its own performance under ICL settings. Within the BERT family, PubMedBERT performs best under SFT (0.808), reflecting the benefit of domain-specific pretraining when sufficient labeled data is available.

Under ICL settings, GPT-5 consistently achieves the highest accuracy across all strategies, exhibiting a significant performance gap relative to other models under both ZS and FS conditions. This result suggests substantial differences in model architecture, training procedures, and pretraining corpora between GPT-5 and the remaining models. However, due to its closed-source architecture and unknown training corpus, it is difficult to determine whether this advantage arises from scale, architecture, or potential exposure to similar training data. In contrast, the other reasoning-oriented models, namely DeepSeek R1 and GPT-o3-mini, do not demonstrate comparable advantages, nor do they show clear improvements over their corresponding general-purpose counterparts, DeepSeek V3.1 and GPT-4o-mini.

Across ICL strategies, all FS variants (F-FS, R-FS, and RAG-FS) consistently outperform the ZS baseline, indicating that even a small number of labeled examples can provide meaningful performance gains. However, no single FS strategy uniformly dominates across models. For most models, except for GPT-5, RAG-FS generally outperforms R-FS, suggesting that semantically similar examples are more informative than randomly selected ones. In some cases, F-FS yields variable performance compared to either R-FS or RAG-FS for each model, implying that example selection is an essential factor that decides the accuracy of FS strategies.

Table 1. Accuracy comparison across different models and settings – Task 1

| Model type | Model | ZS | F-FS | R-FS | RAG-FS | SFT |
|---|---|---|---|---|---|---|
| **BERT-based** | BERT | | | - | | 0.761 |
| | SciBERT | | | - | | 0.789 |
| | BioBERT | | | - | | 0.790 |
| | SBERT | | | - | | 0.780 |
| | RoBERTa | | | - | | 0.774 |
| | RoBERTa-large | | | - | | 0.775 |
| | PubMedBERT | | | - | | <u>0.808</u> |
| **General LLMs** | Llama 3.3-70B | 0.571 | 0.622 | 0.605 | 0.679 | - |
| | GPT-4o-mini | 0.646 | 0.642 | 0.636 | 0.669 | **0.840\*** |
| | DeepSeek V3.1 | <u>0.658 \*</u> | <u>0.704</u> | 0.675 | <u>0.698</u> | - |
| | Qwen3-235B | 0.589 | 0.674 | 0.654 | 0.674 | - |
| | Gemini 2.5 Flash Lite | 0.54 | 0.671 | 0.652 | 0.673 | - |
| **Reasoning LLMs** | GPT-o3-mini | 0.6 | 0.608 | 0.609 | 0.678 | - |
| | GPT-5 | **0.685** | **0.715** | **0.719** | **0.705** | - |
| | DeepSeek R1 | 0.649 | 0.641 | <u>0.681</u> | 0.692 | - |

\* Note: The best results are highlighted in red bold font and second-best results are underlined.

We further examine class-wise precision and recall in Figure 3. Overall, most models exhibit higher precision for the recommended class compared to the low recall for the not recommended class. This pattern indicates that most models struggle to correctly identify articles that experts consider not recommended. The only moderate precision and high recall for the recommended class further suggests that a critical proportion of articles deemed not recommended by experts are nonetheless classified as recommended by LLMs, regardless of the learning strategy employed.

Gemini 2.5 Flash Lite displays a different behavior. Its low precision and high recall for the not recommended class under the ZS setting represents a strong tendency to classify articles as not recommended, including many that experts consider valuable. This contrast highlights a systematic divergence in model bias: Whereas most models appear overly permissive or lenient in their evaluations, Gemini 2.5 Flash Lite adopts an unusually conservative or harsh stance. Even with FS strategies employed, this tendency remains.

Regarding learning strategies, FS approaches (F-FS, R-FS, and RAG-FS) generally improve the identification of not recommended articles, as evidenced by modest increases in recall for this class when moving from ZS to FS settings. However, the slower improvements in precision for the recommended class suggests that the additional articles identified through example-based learning are not always correctly classified. While they partially mitigate the tendency of LLMs toward overly lenient or strict judgments, they do not fundamentally alter the models' underlying class preferences, namely, a general inclination toward positive recommendations for most models and a negative bias for Gemini 2.5 Flash Lite.

Among the FS strategies, the results align closely with the accuracy-based analysis. RAG-FS consistently achieves the top F1-scores across most LLMs, particularly for the not recommended class, indicating a more balanced trade-off between precision and recall. This suggests that incorporating semantically similar examples provides a stable and effective improvement in class-wise performance across models.

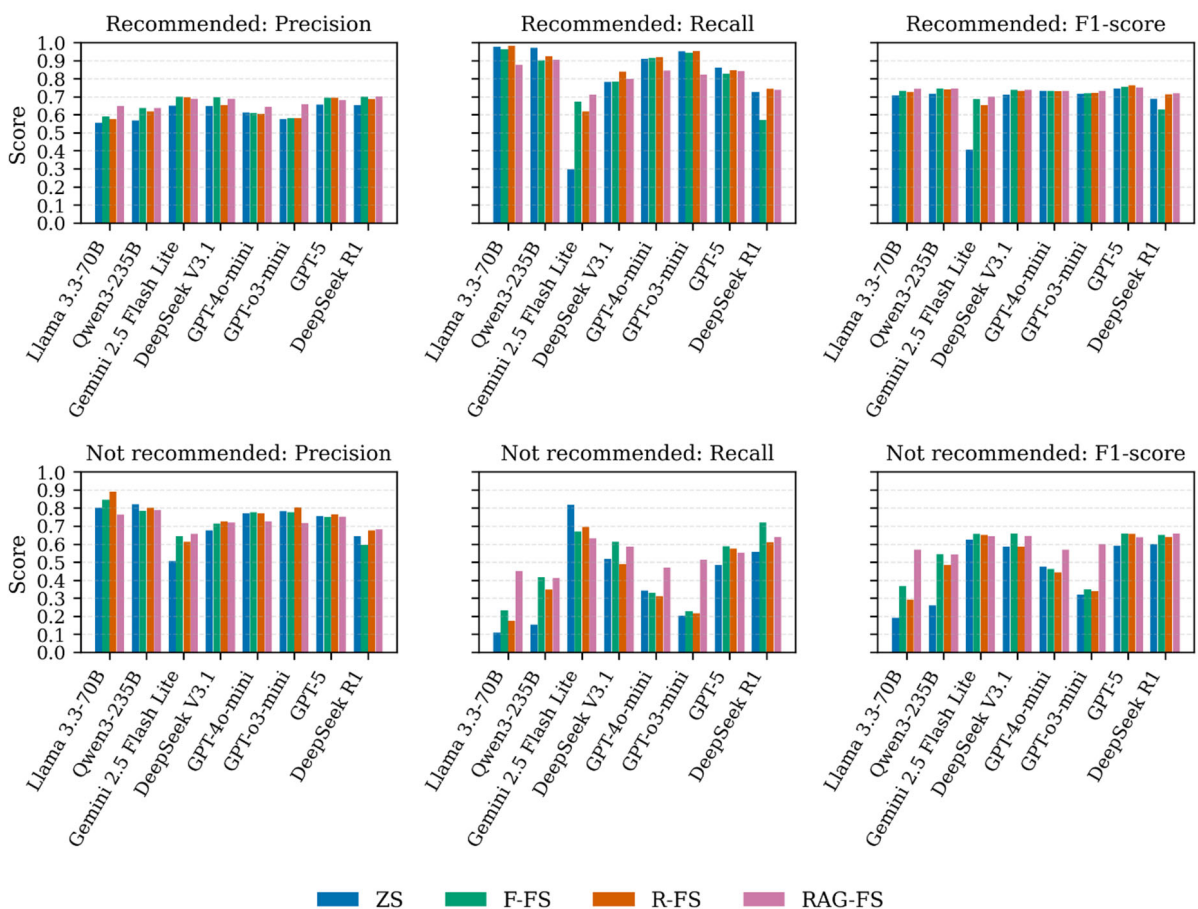

**Figure 3. Model-wise precision and recall comparison with ICL – Task 1**

Figure 4 focuses on the performance of SFT BERT-based and GPT-4o-mini models across the same precision and recall metrics. In contrast to the ICL results, SFT models exhibit uniformly high and stable performance, with relatively small variance across architectures. GPT-4o-mini achieves the strongest overall results, reaching the highest values in precision rate of recommended articles, recall rates of not recommended articles, and F1-scores in both classes. Additionally, precision and recall are much more balanced under SFT for all LLMs.

Compared to general-purpose encoders such as BERT and RoBERTa, domain-adapted models (BioBERT and PubMedBERT) consistently achieve slightly higher recall, suggesting better sensitivity to task-specific semantic signals, confirming their suitability for biomedical and research-evaluation-oriented text classification tasks, which aligns with the biomedical domain of the H1 Connect dataset.

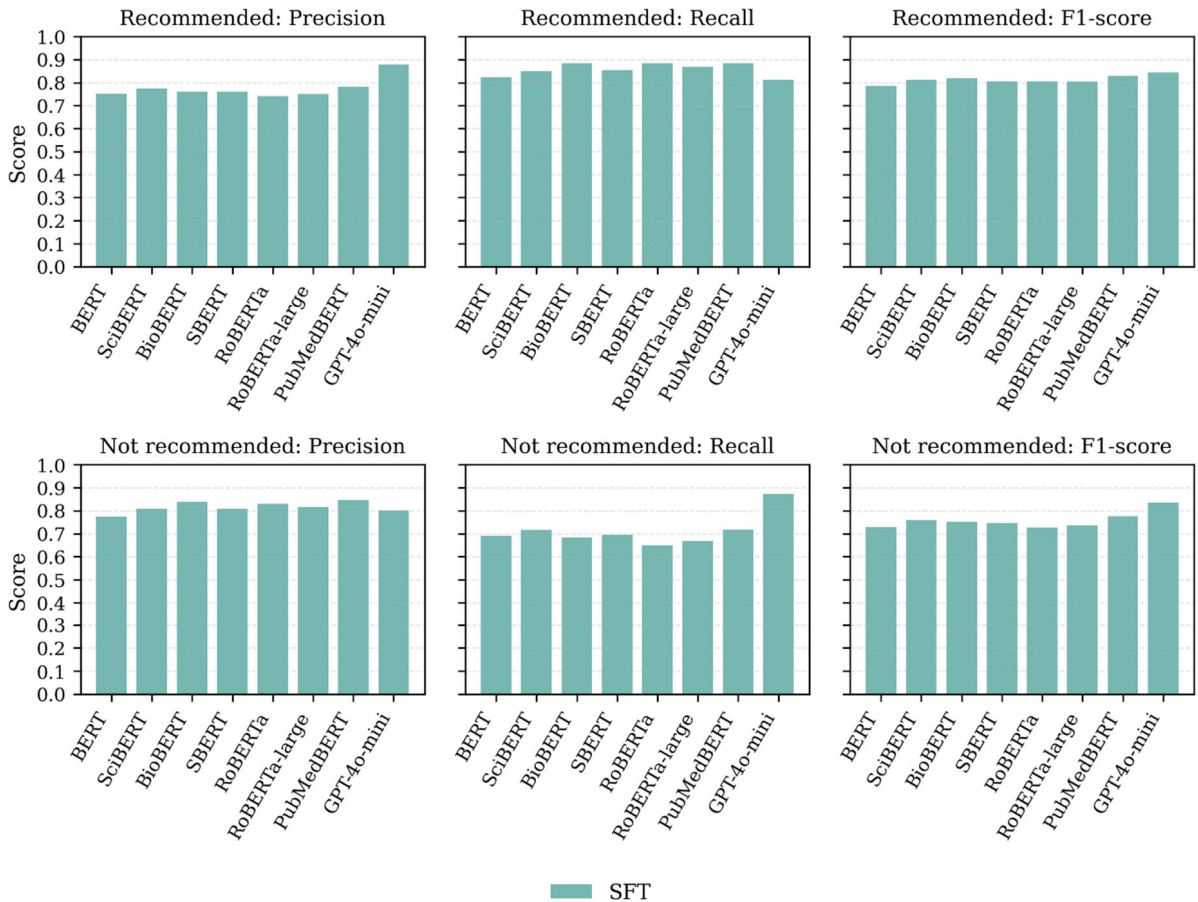

**Figure 4. Model-wise precision and recall comparison with SFT – Task 1**

As a summary for expert alignment results in Task 1, the key findings are as follows: 1) Under ICL settings, most LLMs tend to produce overly positive recommendations, while Gemini 2.5 Flash Lite demonstrates the opposite behavior. 2) FS learning strategies, i.e., providing a few expert-annotated examples to LLMs, can effectively encourage LLMs to give negative opinions but still provide limited improvement to aligning them with expert opinions, RAG-FS comparably is the most effective to accurately identifying both classes. 3) Regardless of learning strategies, GPT-5 exceeds other models significantly on all evaluation metrics when compared against expert opinions. 4) SFT is significantly more effective than any of the ICL strategies, and most of the precision and recall results are much more balanced.

*4.1.2. Citation indicator correlation*

Then, 1,814 out of the 1,907 tested articles in Task 1 were successfully mapped to NCS and CSS values. The distribution of NCS and CSS values is given in Table 2. Figure 5 reports the rank-biserial correlations between Task 1 article-level recommendation scores (binary coded as recommended = 1, not recommended = 0) with the two citation indicators NCS (left orange) and CSS (right blue), both sorting in ascending order. Figure 6 presents the model-wise results of those correlation values. Overall, the results show consistent but moderate positive correlations across models and learning strategies (0.2-0.58).

Table 2. The distribution of NCS and CSS for dataset 1 test set – Task 1

|  | Min | Max | Mean | Std |
|---|---|---|---|---|
| NCS | 0 | 2514.09 | 11.45 | 63.70 |
|  | #0 * | #1 | #2 | #3 |
| CSS | 284 | 429 | 452 | 649 |

* # indicates the number of articles associated with that CSS value.

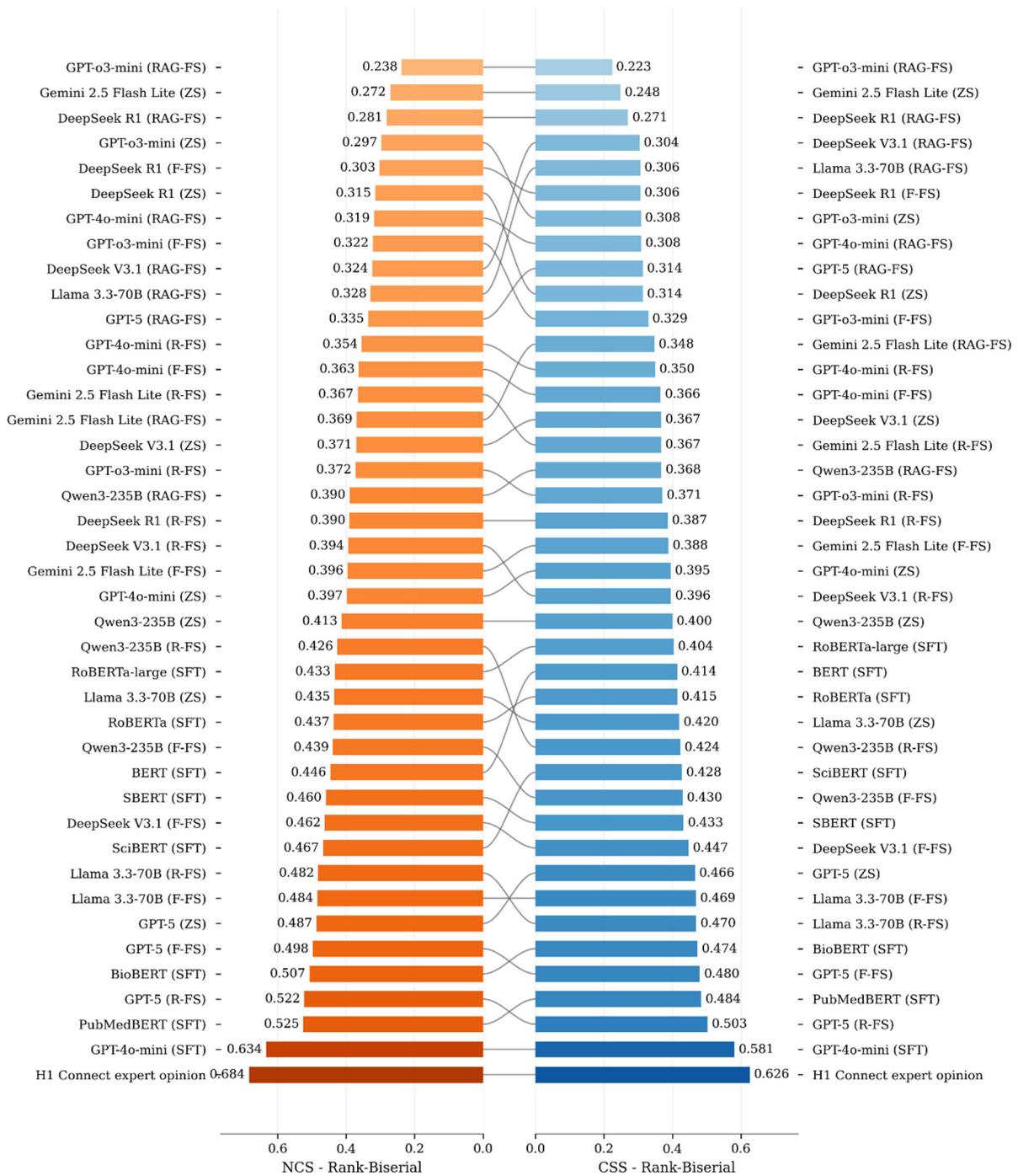

**Figure 5. Rank-biserial correlations between model outputs and NCS & CSS – Task 1**

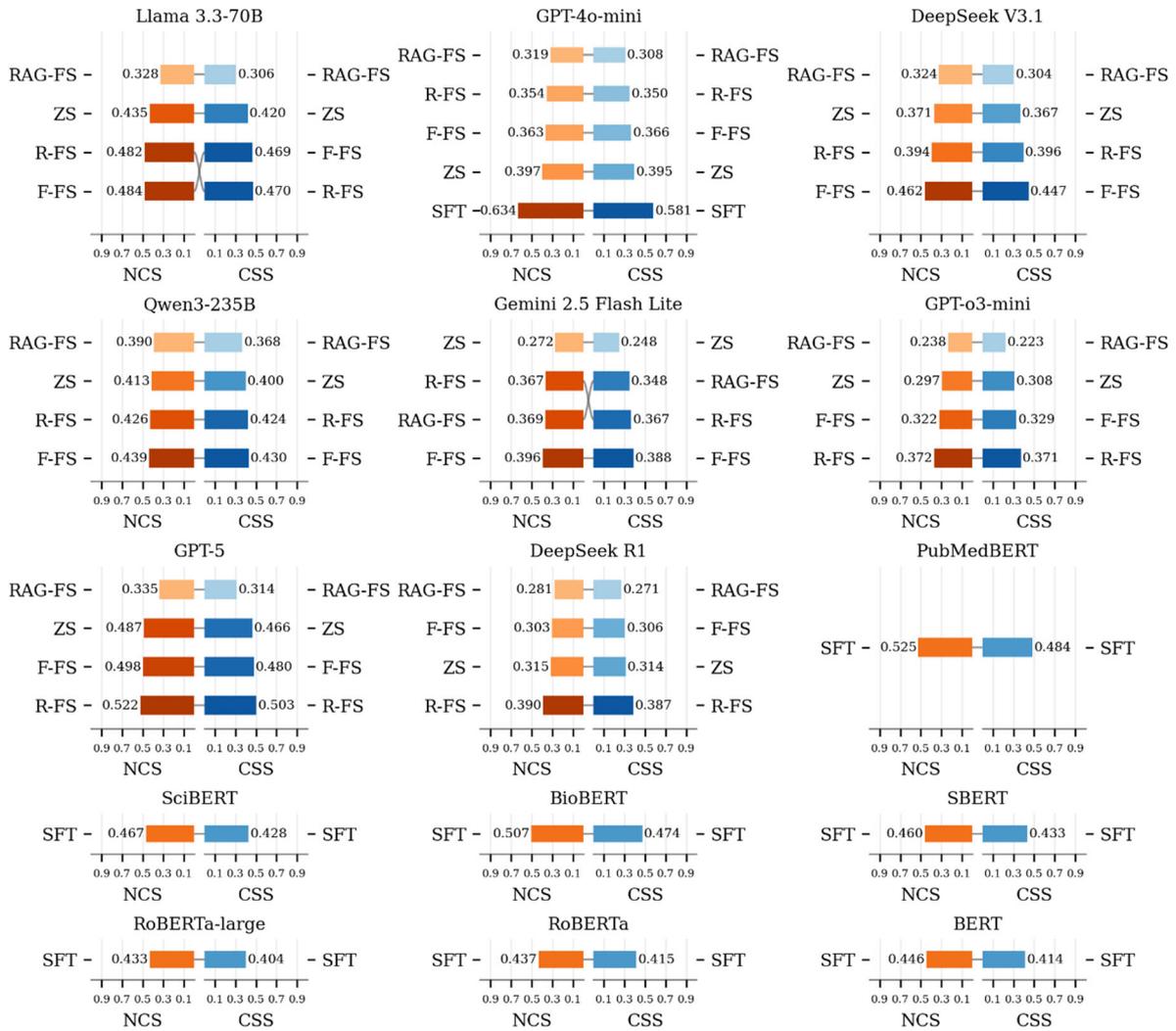

**Figure 6. Model-wise NCS and CSS Rank-biserial correlation comparison – Task 1**

We analyzed the results in Figure 5 and Figure 6 from three perspectives, i.e., model, learning strategy, and citation metrics' differences.

***Which model's outputs present the best correlation with NCS and CSS?***

The highest correlations are observed for the H1 Connect expert opinions, which serves as an reference benchmark. Among the ICL results, GPT-5 exhibits comparatively stronger correlations, achieving values that are competitive with the SFT results of PubMedBERT and BioBERT. In contrast, other LLMs, irrespective of model type, yield relatively weak correlations under ICL settings, with most values clustered below 0.4. Applying SFT leads to a marked improvement in correlation strength across models. Encoder-based SFT models, particularly PubMedBERT and BioBERT, demonstrate notably higher correlations with both CSS and NCS, underscoring the benefit of domain-specific representations when explicit supervision is available.

Across all SFT approaches, GPT-4o-mini under SFT achieves the strongest overall performance, with correlations of approximately 0.63 with CSS and 0.58 with NCS. This represents a substantial improvement over both ICL-based LLMs and fine-tuned encoder baselines, highlighting the effectiveness of supervised adaptation for aligning LLM outputs with citation-based indicators.

***How do different ICL strategies impact the correlation of model outputs with NCS and CSS?***

Overall, ZS settings yield weak correlations with both NCS and CSS, whereas FS strategies generally improve correlation strength relative to ZS. Notably, RAG-FS, despite often achieving higher classification accuracy, does not consistently lead to stronger correlations with citation indicators, particularly for reasoning-oriented models (GPT-o3-mini, GPT-5, and DeepSeek R1). Moreover, RAG-FS does not systematically outperform standard FS approaches, suggesting that incorporating semantically similar examples does not substantially enhance the alignment between model judgments and citation-based metrics.

***What is the difference between NCS and CSS correlations?***

When benchmarked against expert opinions, ICL-based approaches generally exhibit higher correlations with CSS, whereas SFT-based models tend to achieve stronger correlations with NCS. This pattern aligns well with our expectations regarding the granularity of article-level classification. Specifically, ICL appears better suited to reproducing tier-based, qualitative impact judgments that correspond more closely to CSS, while SFT enhances models' sensitivity to fine-grained variations in citation magnitude, resulting in stronger alignment with NCS.

In addition to the standard correlation analysis, we further encoded CSS into binary form (CSS = 0 vs. CSS > 0) to align with the binary recommendation setting and evaluated the corresponding correlations. The results are reported in Supplementary Table 3.

### 4.2. Task 2 - Article rating, coding, and commenting

#### 4.2.1. Human expert alignment - Rating

Table 3 summarizes the overall 3-class rating accuracy (ratings 1, 2, and 3) for Task 2, where models are required to differentiate "Good" (rating 1), "Very Good" (rating 2), and "Exceptional" (rating 3) based on article abstracts. Overall, ICL strategies deliver only moderate accuracy gains and remain far from expert-level consistency for this more nuanced classification problem. Among ICL settings, the best-performing option is GPT-5 with F-FS (0.407), followed by the same model with RAG-FS (0.402) and R-FS (0.399). In contrast, SFT yields the strongest results among the reported settings: Fine-tuned GPT-4o-mini achieves the highest overall accuracy (0.463), outperforming all ICL variants and exceeding the BERT-based SFT baselines. Within the BERT-family SFT models, performance is tightly clustered (approximately 0.446 to 0.460), with BioBERT and RoBERTa achieving the best BERT-level accuracy (0.460) and PubMedBERT close behind (0.459). Compared to Task 1, the gap between BERT-based models with GPT-4o-mini under SFT setting is much smaller. This likely reflects the higher complexity of the task, where SFT still struggles to fully resolve fine-grained rating distinctions.

**Table 3. Accuracy comparison across different models and settings – Task 2**

| Model type | Model | ZS | F-FS | R-FS | RAG-FS | SFT |
|---|---|---|---|---|---|---|
| **BERT-based** | BERT | | | | | 0.447 |
| | SciBERT | | | | | 0.446 |
| | BioBERT | | | - | | <u>0.460</u> |
| | SBERT | | | | | 0.457 |
| | RoBERTa | | | | | <u>0.460</u> |
| | RoBERTa-large | | | | | 0.453 |
| | PubMedBERT | | | | | 0.459 |
| **General LLMs** | Llama 3.3-70B | 0.329 | 0.325 | 0.326 | 0.351 | - |
| | GPT-4o-mini | 0.363 | 0.332 | 0.347 | 0.375 | **0.463** |
| | DeepSeek V3.1 | 0.338 | 0.339 | 0.365 | 0.377 | - |
| | Qwen3-235B | 0.34 | <u>0.390</u> * | 0.374 | 0.368 | - |
| | Gemini 2.5 Flash Lite | 0.36 | 0.367 | 0.361 | <u>0.384</u> | - |
| **Reasoning LLMs** | GPT-o3-mini | 0.367 | 0.385 | <u>0.377</u> | 0.382 | - |
| | GPT-5 | **0.388** * | **0.407** | **0.399** | **0.402** | - |
| | DeepSeek R1 | <u>0.368</u> | 0.384 | 0.373 | 0.357 | - |

* Note. Best results are shown in red bold and second-best results are underlined.

Figure 7 decomposes Task 2 performance into class-wise precision and recall values for three ratings. A consistent pattern is that several model–setting pairs exhibit high recall for a single class but low precision, indicating over-assignment of one rating when uncertainty is high. For example, the recall of rating 2 approaches nearly perfect (Llama 3.3-70B ZS, F-FS, and R-FS) in at least one configuration, while the precision for rating 2 remains near the class prior level for most models, implying that predicting the middle rating captures many true cases but does not reliably discriminate them. Similarly, the precision or recall of rating 3 peaks at 0.992 under specific model–strategy combinations, yet these peaks are not matched by corresponding recall or precision improvements, suggesting degenerate behaviors where the model collapses toward rating 1 or rating 3 for large percentages of inputs. Across models, precision values for rating 2 are relatively stable (often near one third), while the recall distribution varies sharply by setting, which indicates that the main driver of performance differences in ICL is not fine-grained boundary learning but rather shifts in the model's default rating preference induced by prompting together with example selection.

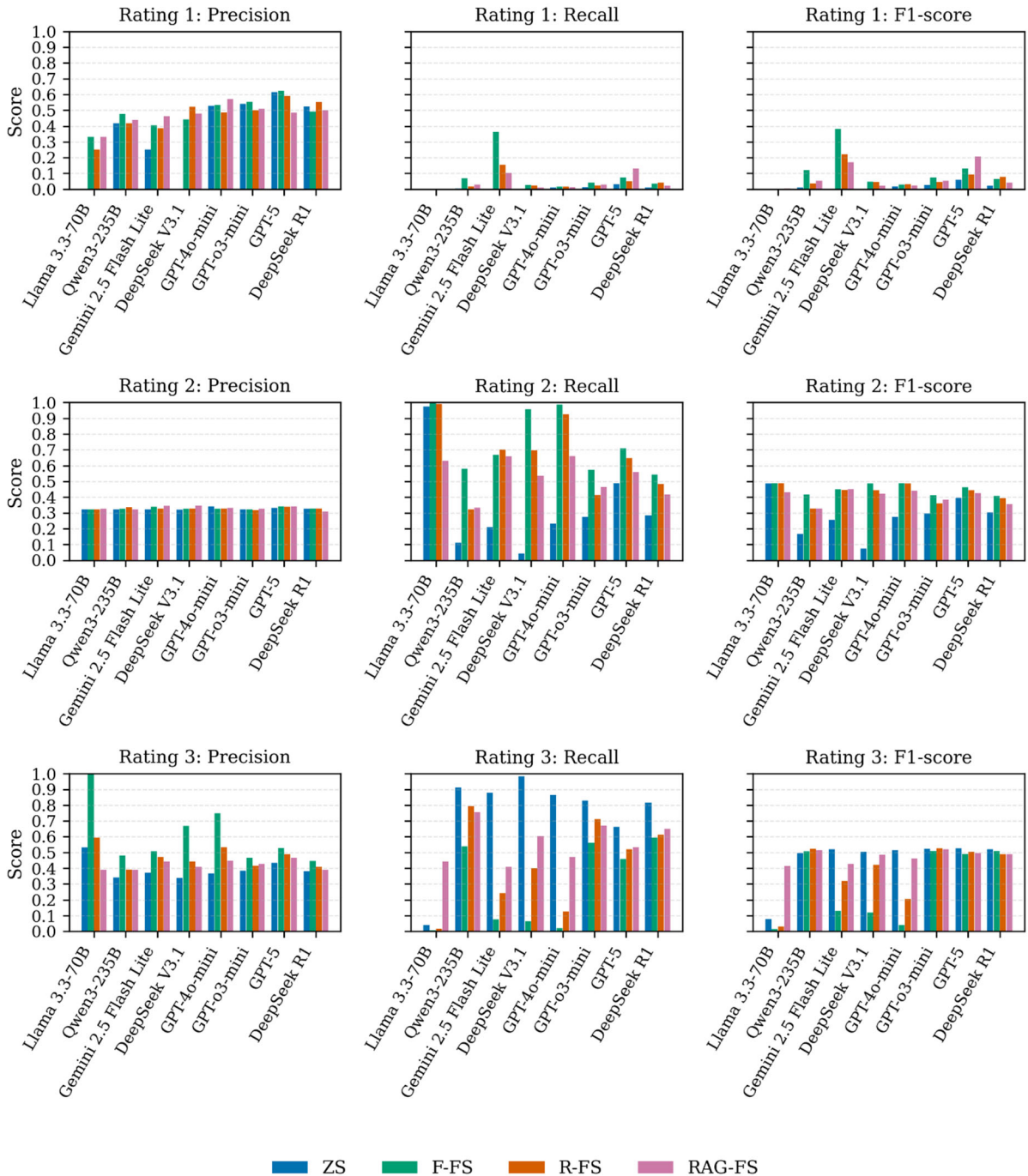

**Figure 7. Model-wise precision and recall comparison with ICL – Task 2**

Figure 8 focuses on SFT models and contrasts their class-wise precision and recall across BERT, SciBERT, BioBERT, RoBERTa, RoBERTa-large, SBERT, PubMedBERT, and the fine-tuned GPT-4o-mini baseline. Overall, SFT produces more coherent class-level behavior than ICL, but meaningful asymmetries remain across ratings. The highest precision of rating 1 is achieved by GPT-4o-mini (0.533), while the highest recall of rating 1 is achieved by SciBERT (0.720), indicating that GPT-4o-mini is more conservative when it predicts rating 1, whereas SciBERT captures a larger fraction of true rating 1 articles. For rating 2, precision of rating 2 articles peaks at 0.361 (RoBERTa), yet recall rate of rating 2 articles remains relatively low and peaks at 0.395 (GPT-4o-mini), suggesting that even after fine-tuning, models still have difficulty recovering rating 2 articles consistently without sacrificing precision. For rating 3, the best results are split: The precision rate of rating 3 articles peaks at 0.583 (SciBERT) while the recall rate of rating 3 articles peaks at 0.526 (GPT-4o-mini), reflecting a persistent challenge in identifying "Exceptional" articles where evidence in abstracts is subtle and sparse. In summary, these findings suggest that SFT substantially stabilizes per-class decision profiles, but the remaining precision–recall imbalance across ratings implies that Task 2 requires either richer input signals (beyond abstracts) or more targeted strategies for learning boundaries between adjacent quality levels. While on this task, the GPT-4o-mini model with SFT does not present significant advancement to other models.

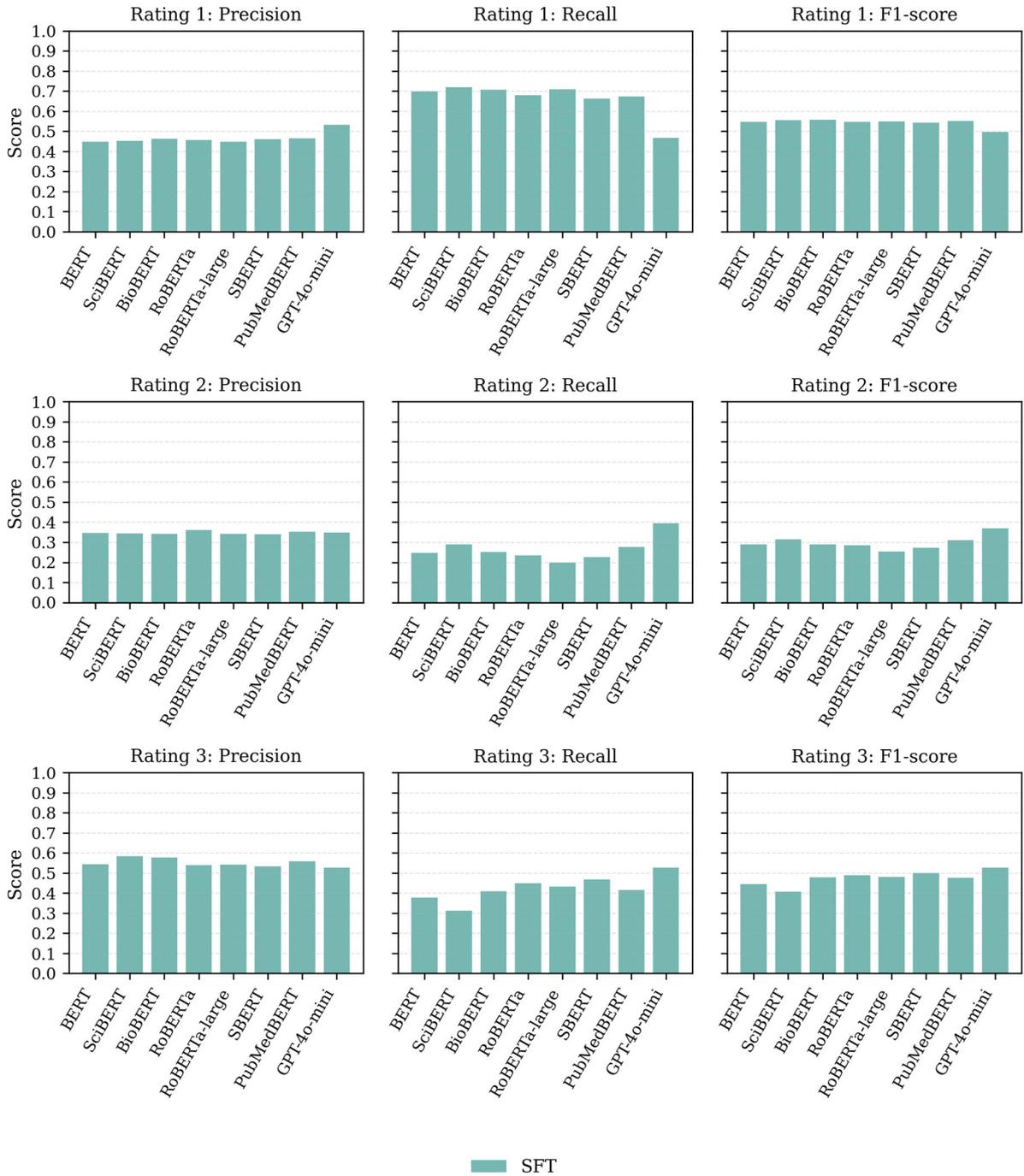

**Figure 8. Model-wise precision and recall comparison with SFT – Task 2**

*4.2.2. Citation indicator correlation*

2,995 of 3,000 test articles in Task 2 were successfully mapped to valid NCS and CSS values. The distribution of NCS and CSS values is given in Table 4. Following the similar approach as Task 1, we calculated the Spearman correlation of model outputs with NCS and CSS, taking each rating 1, 2, and 3 as ordinal categorical variables that represent the quality of articles. The results of Spearman correlation in descending order are given in Figure 9. Figure 10 presents model-wise performance across different learning strategies.

Table 4. The distribution of NCS and CSS for dataset 2 test set – Task 2

|  | Min | Max | Mean | Std |
|---|---|---|---|---|
| NCS | 0 | 1,020.10 | 7.18 | 24.10 |
| CSS | #0 * | #1 | #2 | #3 |
|  | 423 | 999 | 842 | 731 |

* # indicates the number of articles associated with that CSS value.

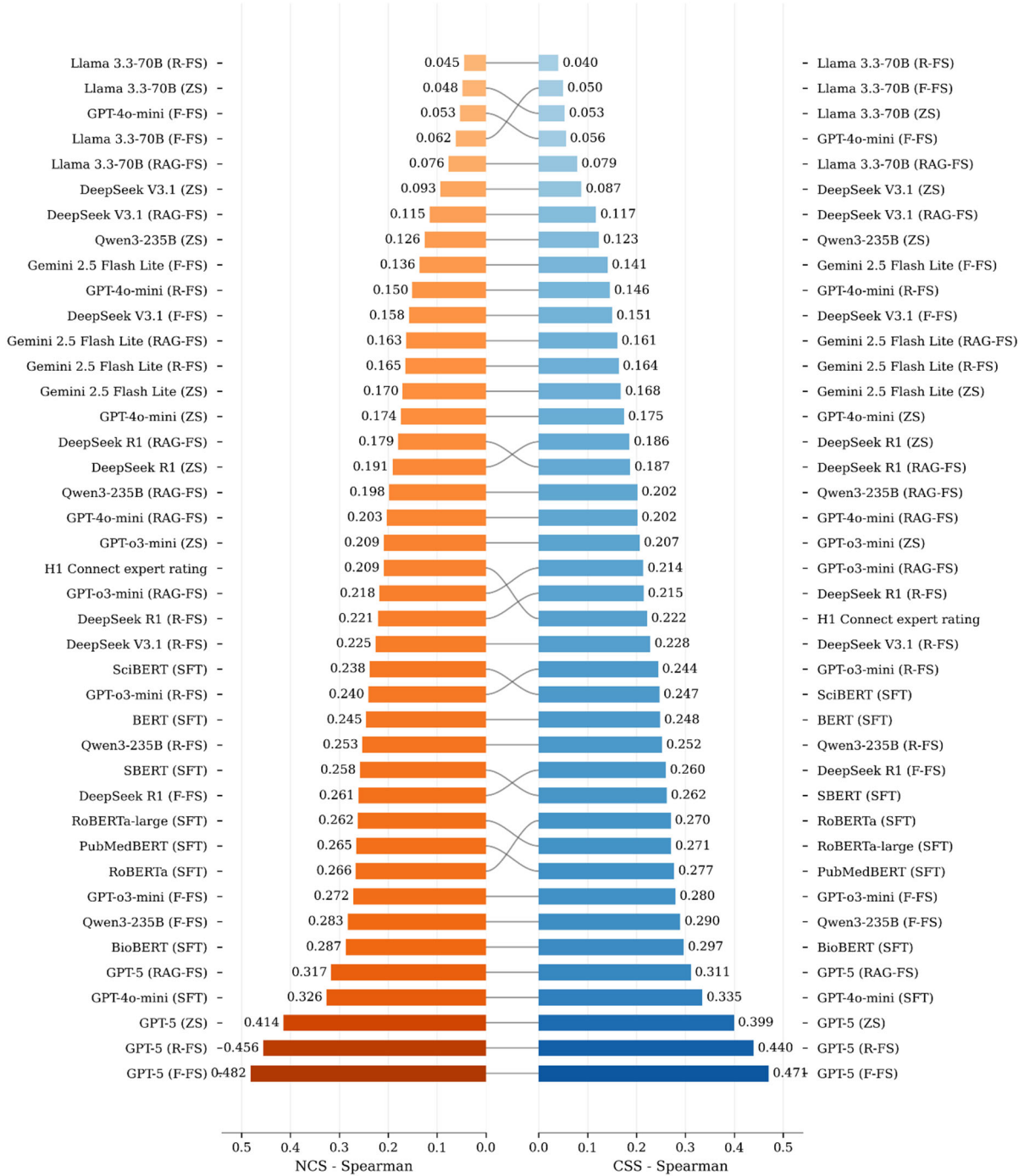

Figure 9. NCS and CSS Spearman correlation comparison – Task 2

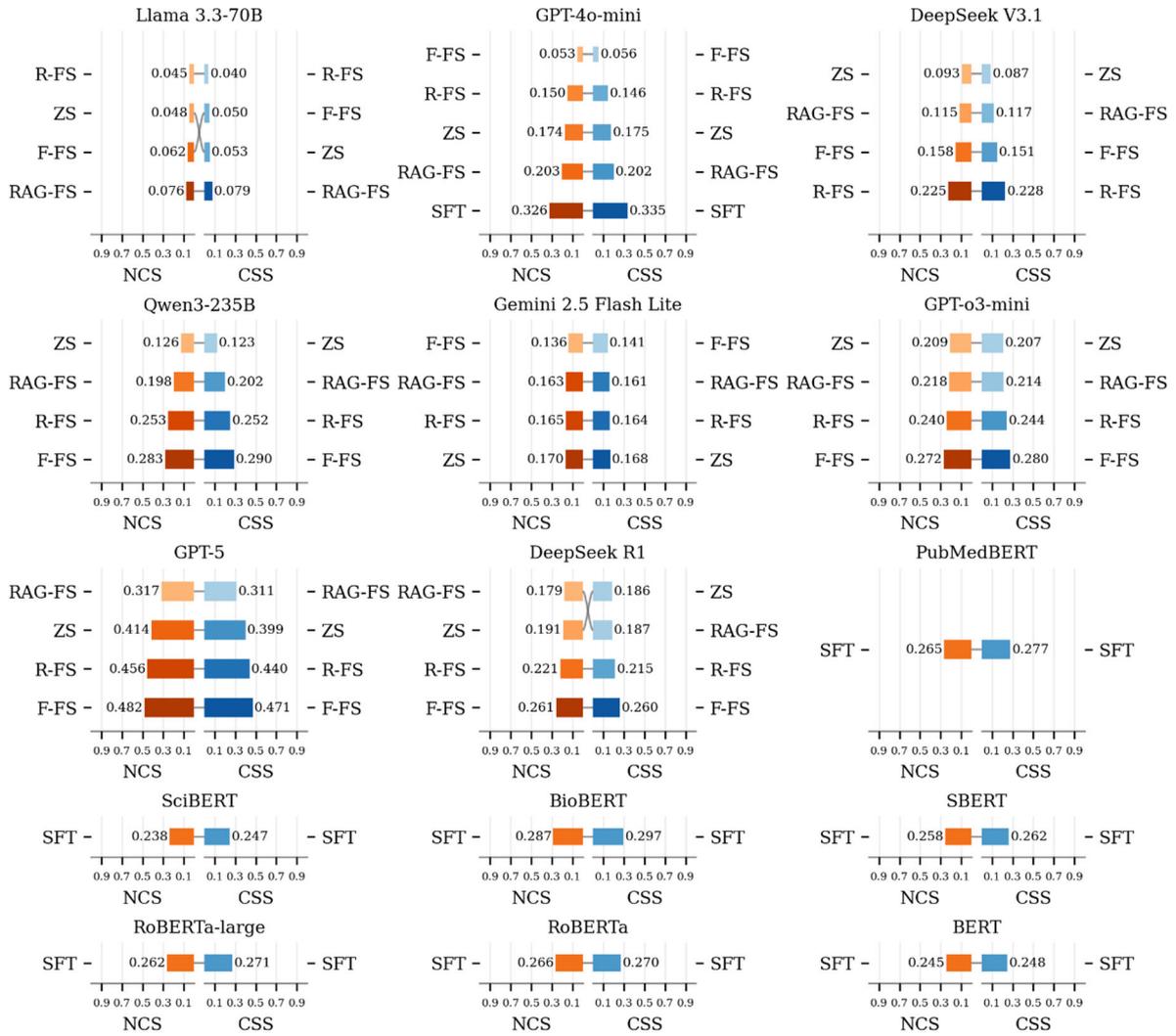

**Figure 10. Model-wise NCS and CSS Spearman correlation comparison – Task 2**

*Which model's outputs present the best correlation with NCS and CSS?*

First, expert ratings do not exhibit the strongest correlations with NCS or CSS for Task 2. This observation raises a fundamental question: When assessing nuanced differences in article quality, it remains unclear which metric or indicator can serve as the most appropriate proxy. Overall, the observed correlations are moderate at best, underscoring the inherent divergence between expert-driven assessments and citation-based measures.

Among all evaluated configurations, GPT-5 demonstrates the strongest alignment with citation indicators, with the FS setting achieving the highest Spearman correlations for both NCS and CSS, followed closely by the R-FS and ZS variants. In contrast, most open-source generative models and smaller LLMs exhibit correlations below 0.25, with minimal differentiation across ICL strategies. SFT of BERT-based models results in mid-range correlations, outperforming most ICL configurations but remaining clearly below GPT-5. Additionally, the results indicate that reasoning-oriented LLMs generally achieve stronger correlations than general-purpose LLMs, suggesting an advantage in capturing evaluation signals related to scholarly impact.

*How do different ICL strategies impact the correlation of model outputs with NCS and CSS?*

For most LLMs, RAG-FS does not emerge as the most effective strategy and, in some cases, negatively affects performance. This effect is particularly pronounced for reasoning-oriented models such as GPT-o3-mini, GPT-5, and DeepSeek R1, which is consistent with the findings observed in Task 1. Moreover, the overall correlation levels in this task are notably lower than those reported for Task 1, indicating a generally weaker alignment between model outputs and citation-based indicators.

Across models, standard FS strategies, including R-FS and F-FS, more consistently yield higher correlations with NCS or CSS. These results suggest that, in this task setting, conventional FS prompting provides more stable benefits than RAG-FS for improving citation-aligned performance.

*What is the difference between NCS and CSS correlations?*

The correlation patterns for NCS and CSS in Task 2 differ from those observed in Task 1. In Task 2, SFT yields stronger correlations with CSS than with NCS, whereas under ICL settings, correlations with both citation indicators remain comparably low. This divergence can be attributed to the increased granularity of Task 2, where expert judgments capture more nuanced distinctions in article quality that are only weakly associated with citation-based metrics. As a result, both expert evaluations and citation indicators exhibit a looser correspondence in this task, leading to uniformly lower correlation values.

*How do LLM outputs perform when benchmarked against NCS and CSS using classification-based evaluation?*

Given that CSS is categorized into four ordinal levels, which correspond closely to our three rating categories (with articles assigned CSS = 0 treated as not recommended, i.e., rating 0), it is also informative to benchmark model outputs against CSS and NCS as an additional ground-truth dimension serving as a proxy for article quality. Accordingly, we retained 2,572 articles with CSS greater than 0 and calculated the classification accuracy of each model's output.

For NCS, we first sorted all 2,995 articles in descending order by their NCS values. The top 990 articles (matching the number of articles assigned rating 3 in the mapped dataset) were labeled as the true class for NCS rating 3. The next 969 articles (corresponding to the number assigned rating 2) were labeled as the true class for NCS rating 2, and the remaining articles were labeled as the true class for NCS rating 1. Under this setting, the classification accuracy of the model outputs and the H1 Connect expert ratings are presented in Figure 11 and Figure 12. Similar to Task 1, in addition to accuracy, we report the correlations among encoded CSS, model outputs, and expert ratings in Supplementary Table 3.

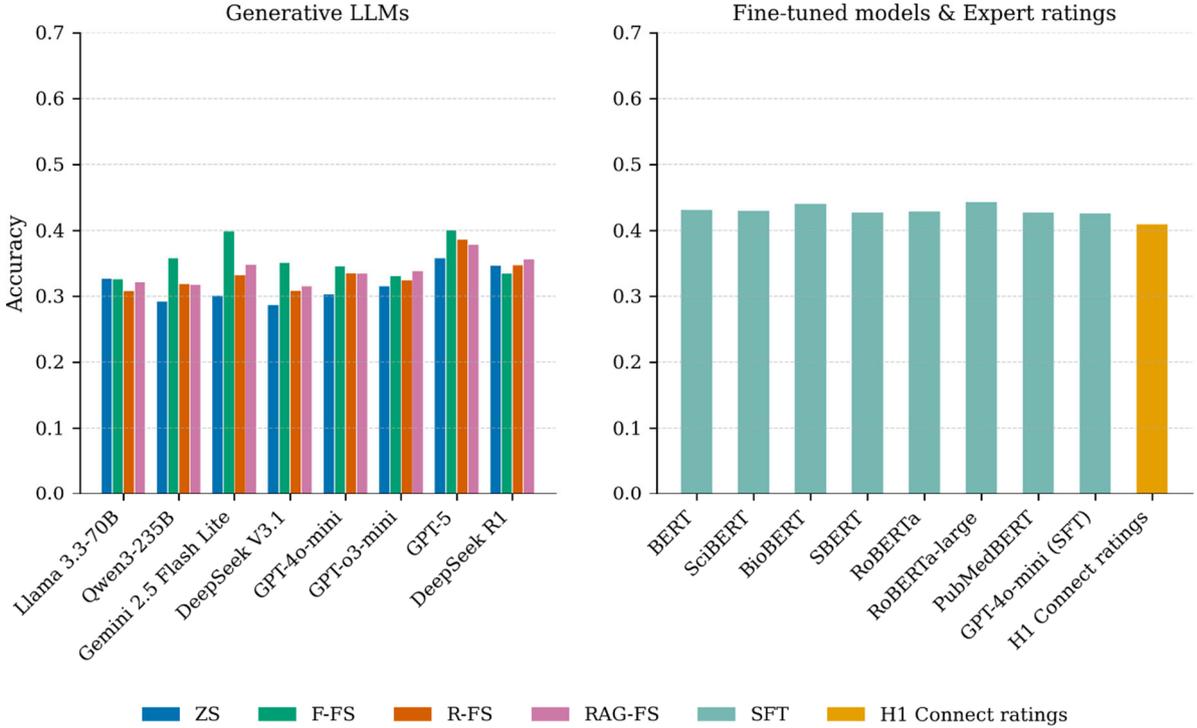

**Figure 11. CSS accuracy compared against CSS baseline – Task 2**

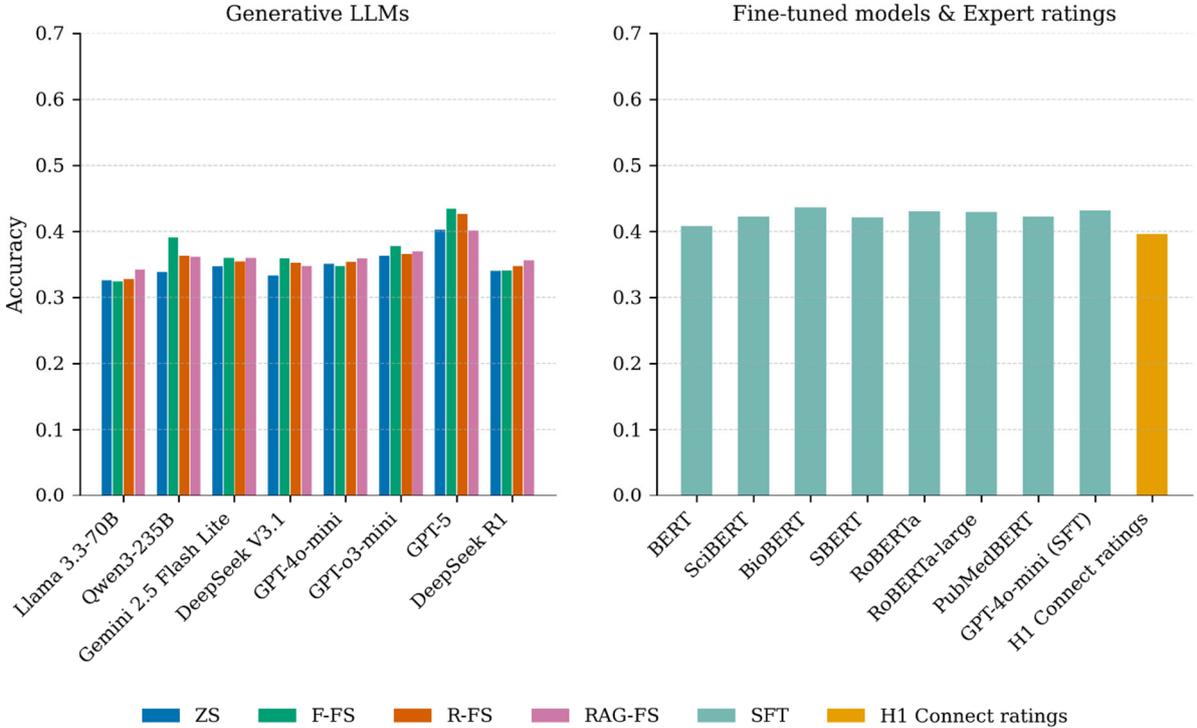

**Figure 12. NCS accuracy compared against NCS baseline – Task 2**

The model comparison results are largely consistent with the correlation analysis. GPT-5 demonstrates the strongest overall performance, while the fine-tuned models achieve higher classification accuracy and exhibit stronger alignment with both citation indicators. Gemini 2.5 Flash-Lite also attains relatively high accuracy in the CSS evaluation (Figure 11). However, this result is largely driven by its tendency to assign the lowest rating (1) to most articles. Because articles with a CSS value of 1 constitute the largest group in the dataset, this behavior artificially inflates the overall accuracy. A more informative assessment would require additional metrics such as precision and recall.

In contrast, GPT-5 and the fine-tuned models generate more balanced predictions across the rating categories. Interestingly, when benchmarked against CSS and NCS, the accuracy of the H1 Connect expert ratings is only around 0.4. This finding suggests a moderate level of alignment between these two widely used indicators of research quality.

*4.2.3. Expert alignment – merit codes and comments*

This analysis is designed to assess the extent to which LLMs can assign expert-style merit codes and generate evaluative comments that resemble those provided by experts. For evaluation, we measure the similarity between LLM- and expert-generated outputs using multiple complementary indicators, including the Jaccard coefficient over classification codes, recall of classification codes, and textual similarity metrics for comments, namely ROUGE-1, ROUGE-2, ROUGE-L, and BLEU.

Table 5 reports the Jaccard coefficients between classification codes assigned by LLMs and those provided by experts. Overall, FS strategies substantially improve code overlap compared to ZS prompting across all evaluated models. Both F-FS and R-FS settings yield consistent gains, while RAG-FS learning achieves the highest Jaccard scores for most models. This pattern suggests that exposing models to expert-labeled examples enhances their ability to recover core evaluative dimensions, even when multiple codes may apply to a single article.

Among all the models, Gemini 2.5 Flash Lite demonstrates the strongest overall performance under all settings, with Jaccard coefficients exceeding those of other generative models. However, SFT on GPT-4o-mini does not universally outperform in-context learning for Task 2, indicating that code assignment benefits more from exemplar exposure than from parameter updates alone. These findings highlight that merit codes represent a multi-label, interpretive task where contextual grounding plays a critical role.

Table 5. Classification code Jaccard Coefficient across LLMs – Task 2

| Model type | Model | ZS | F-FS | R-FS | RAG-FS | SFT |
|---|---|---|---|---|---|---|
| General LLMs | Llama 3.3-70B | 0.347 | 0.413 | 0.404 | 0.424 | - |
| | GPT-4o-mini | 0.263 | 0.438 | 0.434 | 0.448 | 0.397 |
| | DeepSeek V3.1 | 0.334 | <u>0.467</u> | 0.448 | 0.457 | - |
| | Qwen3-235B | 0.333 | 0.367 | 0.385 | 0.408 | - |
| | Gemini 2.5 Flash Lite | **0.452*** | **0.503** | **0.463** | **0.478** | - |
| Reasoning LLMs | GPT-o3-mini | <u>0.385</u> | 0.446 | <u>0.461</u> | <u>0.463</u> | - |
| | GPT-5 | 0.291 | 0.405 | 0.392 | 0.4 | - |
| | DeepSeek R1 | 0.312 | 0.42 | 0.41 | 0.424 | - |

* Note. Best results are shown in red bold and second-best results are underlined.

Table 6 presents code-wise F1 scores, indicating the extent to which LLMs can recover expert-identified article merits. Across models and strategies, F1-score is consistently higher for frequently assigned codes such as *New Finding (A)* and *Hypothesis (B)*, while more specialized or evaluative codes like *Controversial (H)* and *Refutation (I)* exhibit lower F1-scores. FS strategies again improve recall across most codes, still suggesting that LLMs benefit from explicit demonstrations of how experts operationalize evaluative criteria. Importantly, lower F1-scores for rare codes reflect both class imbalance and the inherent difficulty of inferring nuanced judgments from abstracts alone.

Table 6. Code-wise F1-score for merit code assignment – Task 2

| Model | LS | A | B | C | D | E | F | G | H | I | J | K |
|---|---|---|---|---|---|---|---|---|---|---|---|---|
| Llama 3.3-70B | ZS | 0.853 | 0.046 | 0.21 | 0.398 | 0.281 | 0.281 | 0.588 | 0.198 | 0.247 | 0.056 | 0.096 |
| | R-FS | 0.858 | 0.259 | 0.208 | 0.459 | 0.284 | 0.13 | 0.484 | 0.145 | 0.278 | 0.085 | 0.205 |
| | F-FS | 0.868 | 0.238 | 0.21 | 0.451 | 0.29 | 0.161 | 0.513 | 0.143 | 0.242 | 0.095 | 0.225 |
| | RAG-FS | 0.861 | 0.333 | 0.334 | 0.463 | 0.297 | 0.336 | 0.543 | 0.148 | 0.214 | 0.147 | 0.197 |
| GPT-4o-mini | ZS | 0.7 | 0.021 | 0.309 | 0.472 | 0.314 | 0.225 | 0.134 | 0.2 | 0.001 | 0.019 | 0.2 |
| | R-FS | 0.849 | 0.265 | 0.168 | 0.478 | 0.284 | 0.315 | 0.621 | 0.167 | 0.001 | 0.273 | 0.286 |
| | F-FS | 0.86 | 0.196 | 0.213 | 0.399 | 0.267 | 0.311 | 0.713 | 0.189 | 0.053 | 0.333 | 0.25 |
| | RAG-FS | 0.85 | 0.326 | 0.297 | 0.51 | 0.299 | 0.343 | 0.641 | 0.186 | 0.095 | 0.25 | 0.167 |
| | SFT | 0.852 | 0.205 | **0.442** | 0.485 | 0.164 | 0.146 | 0.749 | 0.063 | 0.046 | 0.25 | 0.19 |
| DeepSeek V3.1 | ZS | 0.776 | 0.096 | 0.346 | 0.363 | 0.286 | 0.286 | 0.687 | 0.16 | 0.224 | 0.208 | 0.235 |
| | R-FS | 0.871 | 0.275 | 0.372 | 0.449 | 0.32 | 0.341 | 0.746 | 0.183 | 0.234 | 0.261 | 0.235 |
| | F-FS | 0.872 | 0.214 | 0.36 | 0.467 | 0.332 | **0.408** | 0.721 | 0.138 | 0.255 | 0.311 | 0.272 |
| | RAG-FS | 0.866 | 0.318 | <u>0.424</u> | 0.464 | 0.334 | 0.366 | 0.756 | <u>0.218</u> | 0.232 | 0.317 | 0.128 |
| | ZS | 0.856 | 0.322 | 0.385 | 0.377 | 0.239 | 0.22 | 0.639 | 0.149 | 0.19 | 0.062 | 0.205 |

| Model | LS | A | B | C | D | E | F | G | H | I | J | K |
|---|---|---|---|---|---|---|---|---|---|---|---|---|
| Qwen3-235B | R-FS | 0.862 | 0.344 | 0.356 | 0.452 | 0.288 | 0.281 | 0.688 | 0.169 | 0.196 | 0.123 | 0.222 |
| | F-FS | 0.858 | 0.297 | 0.37 | 0.39 | 0.238 | 0.27 | 0.633 | 0.133 | 0.272 | 0.08 | 0.171 |
| | RAG-FS | 0.861 | 0.369 | 0.381 | 0.478 | 0.316 | 0.313 | 0.734 | 0.2 | 0.217 | 0.162 | 0.246 |
| Gemini 2.5 Flash Lite | ZS | 0.847 | 0.063 | 0.128 | 0.505 | 0.133 | 0.281 | 0.726 | 0.01 | 0.188 | 0.227 | 0.462 |
| | R-FS | 0.86 | 0.18 | 0.21 | 0.511 | 0.237 | 0.329 | 0.71 | 0.067 | 0.289 | 0.178 | 0.297 |
| | F-FS | 0.864 | 0.124 | 0.225 | 0.519 | 0.127 | 0.356 | 0.644 | 0.019 | 0.226 | 0.257 | 0.318 |
| | RAG-FS | 0.859 | 0.272 | 0.247 | 0.514 | 0.226 | 0.332 | 0.714 | 0.075 | 0.226 | 0.229 | 0.3 |
| GPT-o3-mini | ZS | 0.851 | 0.067 | 0.337 | 0.356 | 0.27 | 0.357 | 0.726 | 0.126 | 0.167 | 0.255 | 0.261 |
| | R-FS | 0.869 | 0.171 | 0.316 | 0.444 | 0.278 | 0.355 | 0.768 | 0.152 | 0.241 | 0.306 | 0.271 |
| | F-FS | 0.867 | 0.162 | 0.381 | 0.409 | 0.277 | 0.346 | 0.686 | 0.163 | 0.178 | 0.33 | 0.222 |
| | RAG-FS | 0.867 | 0.211 | 0.347 | 0.443 | 0.259 | 0.38 | 0.757 | 0.162 | 0.231 | 0.313 | 0.216 |
| GPT-5 | ZS | 0.782 | 0.183 | 0.31 | 0.457 | 0.343 | 0.377 | 0.707 | 0.206 | 0.168 | 0.274 | 0.224 |
| | R-FS | 0.873 | 0.289 | 0.334 | 0.554 | 0.332 | 0.401 | 0.77 | 0.214 | 0.169 | 0.353 | 0.213 |
| | F-FS | 0.876 | 0.266 | 0.348 | 0.557 | 0.347 | 0.404 | 0.768 | 0.215 | 0.166 | 0.33 | 0.213 |
| | RAG-FS | 0.871 | 0.325 | 0.349 | 0.561 | 0.35 | 0.397 | 0.772 | 0.226 | 0.175 | 0.344 | 0.227 |
| DeepSeek R1 | ZS | 0.761 | 0.147 | 0.277 | 0.416 | 0.278 | 0.253 | 0.617 | 0.173 | 0.164 | 0.15 | 0.174 |
| | R-FS | 0.866 | 0.295 | 0.331 | 0.45 | 0.323 | 0.313 | 0.689 | 0.173 | 0.285 | 0.298 | 0.275 |
| | F-FS | 0.866 | 0.264 | 0.401 | 0.442 | 0.321 | 0.337 | 0.739 | 0.169 | 0.219 | 0.308 | 0.222 |
| | RAG-FS | 0.864 | 0.334 | 0.386 | 0.493 | 0.324 | 0.365 | 0.71 | 0.182 | 0.147 | 0.294 | 0.103 |

\* Note: Best results are shown in red bold and second-best results are underlined.

A - New Finding    B – Hypothesis    C – Good for Teaching    D – Technical Advance

E – Confirmation    F – Novel Drug Target    G – Review    H – Controversial

I – Refutation    J – Systematic Review    K - Negative

Table 7 evaluates the similarity between model-generated comments and expert comments using standard text similarity metrics. Overall similarity scores remain moderate, indicating partial alignment in content but substantial variation in phrasing and emphasis. FS strategies consistently outperform ZS prompting, reinforcing the importance of exemplary-based grounding for generative commentary. Nevertheless, even under the best-performing configurations, model-generated comments should be interpreted as approximations of expert reasoning rather than substitutes for expert discourse.

**Table 7. ROGUE and BLEU scores between LLM and expert comments – Task 2**

| Model | Method | ROUGE-1 | ROUGE-2 | ROUGE-L | BLEU |
|---|---|---|---|---|---|
| Llama 3.3-70B | ZS | 0.358 | 0.085 | 0.186 | 0.020 |
| | R-FS | 0.370 | 0.098 | 0.197 | 0.023 |
| | F-FS | 0.371 | 0.100 | 0.200 | 0.023 |
| | RAG-FS | 0.371 | 0.098 | 0.196 | 0.023 |
| GPT-4o-mini | ZS | 0.324 | 0.064 | 0.166 | 0.012 |
| | R-FS | 0.340 | 0.075 | 0.176 | 0.015 |
| | F-FS | 0.339 | 0.075 | 0.176 | 0.015 |
| | RAG-FS | 0.343 | 0.076 | 0.177 | 0.016 |
| | SFT | 0.349 | 0.105 | 0.202 | 0.033 |
| DeepSeek V3.1 | ZS | 0.315 | 0.059 | 0.157 | 0.010 |
| | R-FS | 0.331 | 0.068 | 0.166 | 0.013 |
| | F-FS | 0.334 | 0.070 | 0.168 | 0.013 |
| | RAG-FS | 0.335 | 0.071 | 0.168 | 0.013 |
| Qwen3-235B | ZS | 0.338 | 0.069 | 0.161 | 0.013 |
| | R-FS | 0.350 | 0.076 | 0.167 | 0.014 |
| | F-FS | 0.348 | 0.075 | 0.165 | 0.014 |
| | RAG-FS | 0.351 | 0.076 | 0.167 | 0.014 |
| Gemini 2.5 Flash Lite | ZS | 0.325 | 0.069 | 0.171 | 0.013 |
| | R-FS | 0.333 | 0.074 | 0.176 | 0.014 |
| | F-FS | 0.333 | 0.075 | 0.177 | 0.014 |
| | RAG-FS | 0.334 | 0.075 | 0.177 | 0.014 |
| GPT-o3-mini | ZS | 0.272 | 0.053 | 0.151 | 0.009 |
| | R-FS | 0.281 | 0.059 | 0.157 | 0.009 |
| | F-FS | 0.281 | 0.059 | 0.157 | 0.009 |
| | RAG-FS | 0.284 | 0.058 | 0.157 | 0.009 |
| GPT-5 | ZS | 0.272 | 0.038 | 0.129 | 0.006 |
| | R-FS | 0.278 | 0.041 | 0.133 | 0.007 |
| | F-FS | 0.279 | 0.042 | 0.134 | 0.007 |
| | RAG-FS | 0.280 | 0.041 | 0.134 | 0.007 |
| DeepSeek R1 | ZS | 0.289 | 0.049 | 0.144 | 0.008 |
| | R-FS | 0.304 | 0.054 | 0.149 | 0.009 |
| | F-FS | 0.303 | 0.055 | 0.150 | 0.009 |
| | RAG-FS | 0.305 | 0.054 | 0.149 | 0.009 |

\* Note. Best results are shown in red bold and second-best results are underlined.

### 4.3. Task 1 and Task 2 - Data Efficiency Analysis

Given that SFT generally demonstrates stronger performance than ICL in recommendation and rating tasks, an important question arises regarding the amount of labeled data required to achieve satisfactory evaluation performance. To further investigate this issue, we conducted an additional grid-based training experiment in which different proportions of the training data were used to fine-tune the models, and the resulting classification accuracy was evaluated. For the BERT-based models, incremental training ratios of 20%, 40%, 60%, and 80% of the dataset were employed, with the remaining data used as the test set. For GPT-4o-mini, only the 20% and 80% training ratios were evaluated due to API cost constraints.

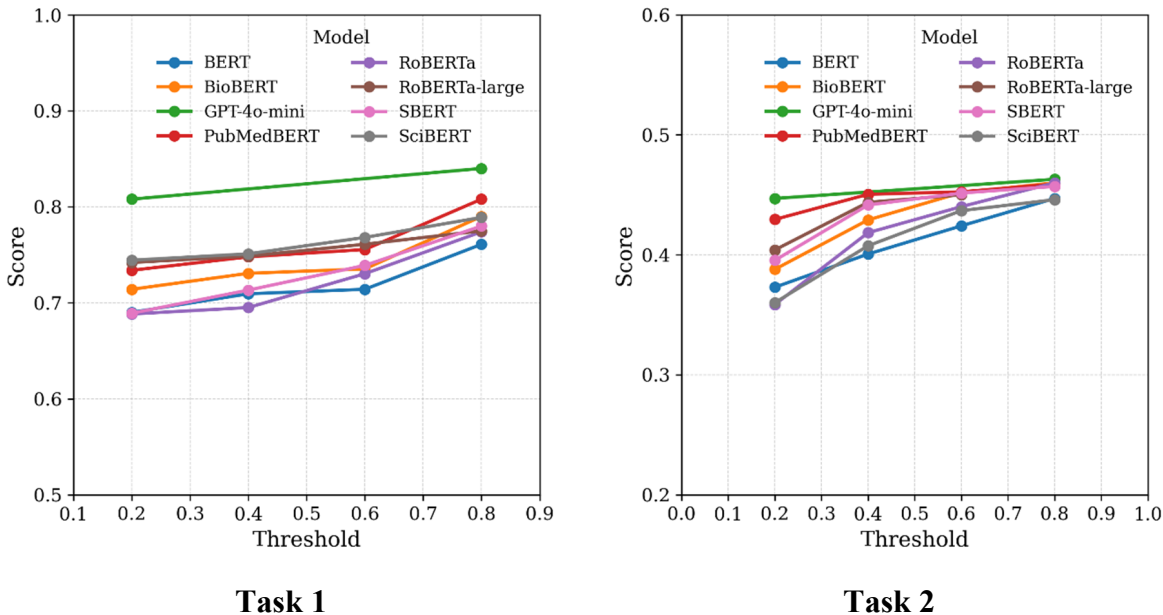

Task 1                                                Task 2

**Figure 13. The impact of training set ratio to accuracies on both tasks**

Results from Figure 13 provide several clear observations. First, model performance consistently improves as the amount of training data increases, indicating that exposure to more labeled examples enhances classification accuracy in both tasks. However, the initial performance levels, the rate of improvement, and the performance gap between the fine-tuned GPT-4o-mini and the BERT-based models differ substantially across the two tasks, which reflects differences in task complexity.

In Task 1, most models already achieve an accuracy above 0.68 when trained with only 20% of the data. In contrast, Task 2 remains considerably more challenging. Even when 80% of the training data is used, the best-performing model still achieves an accuracy below 0.5. Furthermore, the improvement in accuracy as the training data ratio increases is much more modest in Task 2 compared with Task 1. These results suggest that Task 2 requires stronger modeling capability and potentially more informative training signals to achieve comparable performance.

## 5. Conclusions and discussion

In this study, we systematically evaluated the ability of LLMs to support post-publication research evaluation by comparing their output with expert annotations and citation-based indicators. Using two complementary tasks, namely high-quality article identification and fine-grained rating, coding, and commenting, we assessed how different model architectures and learning strategies perform under realistic evaluation conditions.

*RQ1: To what extent can LLMs perform post-publication research evaluation tasks in comparison with expert judgments?*

The results show that LLMs can approximate expert judgments in coarse-grained evaluation tasks but remain limited in nuanced assessment. In Task 1, several models achieve high accuracy in distinguishing recommendations from not recommended articles, particularly under SFT settings. This indicates that LLMs can capture broad quality signals from titles and abstracts. In contrast, Task 2 reveals substantial performance degradation. Models struggle to reliably differentiate among multiple quality levels and to reproduce expert-style ratings, merit codes, and comments. These findings suggest that either abstract-level information is insufficient for resolving fine-grained evaluative distinctions that rely on deeper methodological and contextual reasoning, or the current models are not (yet) capable of distinguishing between different high-quality levels. Overall, LLMs are effective for broad screening but not for detailed expert-level evaluation.

*RQ2: Which LLM and learning strategies yield the most reliable performance for post-publication research evaluation?*

Across both tasks, SFT yields the most reliable and stable performance. Fine-tuned models consistently outperform in-context learning strategies in accuracy, precision–recall balance, and alignment with expert annotations. FS improves performance relative to ZS settings, but the gains are inconsistent across models and tasks. RAG-FS does not consistently outperform standard FS approaches and can even degrade performance for reasoning-oriented models. No single open or fine-tuned model dominates all tasks. Domain-adapted encoder models perform competitively under supervision, while GPT-5 shows strong results but remains difficult to interpret due to its closed-source nature. Overall, learning strategy has a stronger impact on performance than model families.

*RQ3: When benchmarked against citation-based indicators, how do model choice and learning strategies influence performance?*

LLM-based evaluations show moderate correlations with citation-based indicators, reflecting partial alignment with research impact. In Task 1, expert judgments exhibit the strongest correlations with both NCS and CSS, serving as an upper-bound benchmark. SFT improves model alignment with citation indicators, while in-context learning aligns more closely with tier-based impact measures. In Task 2, correlations weaken substantially for both experts and models, highlighting the growing divergence between nuanced expert assessments and citation-based metrics. These results confirm that expert judgment and citation indicators capture related but non-equivalent dimensions of research quality, and that LLMs inherit this divergence rather than resolve it.

This study has several limitations that should be considered. First, the evaluation relies solely on titles and abstracts, which provide only partial information about research quality; important aspects such as methodological rigor and novelty are often reflected only in the full text. Second, the dataset is derived from H1 Connect recommendations within the biomedical domain, which may limit the generalizability of the findings to other disciplines. Third, citation indicators (NCS and CSS) capture scholarly impact rather than intrinsic research quality, which partly explains the moderate correlations observed between model outputs, expert judgments, and citation metrics. Future research could extend this work by incorporating full-text inputs, evaluating multiple disciplinary datasets, exploring additional evaluation benchmarks beyond citation indicators, and investigating hybrid human–AI evaluation workflows.

## Competing interests

The authors declare that they have no competing interests.

## Acknowledgements

A pilot study of this work was accepted by and presented at the 20[th] International Conference on Scientometrics and Informetrics in Yerevan, Armenia. Mengjia Wu and Yi Zhang were supported by the Commonwealth Scientific and Industrial Research Organization (CSIRO), Australia, in conjunction with the National Science Foundation (NSF) of the United States, under CSIRO-NSF #2303037. Mengjia Wu was also supported by Australian Research Council under Discovery Early Career Researcher Award DE260101493.

**Appendix**

**Supplementary Table 1. Task 1 prompts**

| Learning strategy | Prompt |
|---|---|
| ZS | You are an academic expert in the biomedical field, evaluating research articles based on scientific rigor, replicability, data analysis, and study limitations.<br><br>If you recommend this paper, reply with 1. If you do not recommend this paper, reply with 2. Reply only with 1 or 2 with nothing else.<br><br>Now please give recommendations for this paper:<br>[The target article's abstract] * |
| F-FS and R-FS | You are an academic expert in the biomedical field, evaluating research articles based on scientific rigor, replicability, data analysis, and study limitations.<br><br>If you recommend this paper, reply with 1. If you do not recommend this paper, reply with 2. Reply only with 1 or 2 with nothing else.<br>Here are six examples with expert recommendations already given, please refer to them when you make recommendations.<br><br>[Abstract 1]<br>1<br><br>[Abstract 2]<br>1<br><br>[Abstract 3]<br>1<br><br>[Abstract 4]<br>2<br><br>[Abstract 5]<br>2<br><br>[Abstract 6]<br>2<br><br>Now please give recommendations for this paper:<br>[The target article's abstract] |
| RAG-FS | You are an academic expert in the biomedical field, evaluating research articles based on scientific rigor, replicability, data analysis, and study limitations.<br><br>If you recommend this paper, reply with 1. If you do not recommend this paper, reply with 2. Reply only with 1 or 2 with nothing else. |

| Learning strategy | Prompt |
|---|---|
| | Here are the six semantically similar (6-nearest neighbor) examples with their expert recommendations already given. Please refer to them carefully when you make recommendations.<br><br>[Abstract 1]<br>[1/2]<br><br>[Abstract 2]<br>[1/2]<br><br>[Abstract 3]<br>[1/2]<br><br>[Abstract 4]<br>[1/2]<br><br>[Abstract 5]<br>[1/2]<br><br>[Abstract 6]<br>[1/2]<br><br>Now please give recommendations for this paper:<br>[The target article's abstract] |

* Content in [] are template content that varies for each target article.

**Supplementary Table 2. Task 2 prompts**

| Learning strategy | Prompt |
|---|---|
| ZS | You are an expert who provides valuable insights, recommendations, and evaluations related to published research and clinical trial data.<br><br>Specifically, you are responsible for<br>1) Identifying and recommending articles, studies, and research outputs that you consider significant and impactful in the relevant fields.<br>You assess the quality, relevance, and potential impact of the research and provide your expert opinion on its importance.<br>2) evaluating clinical trial results and providing insights into their implications for clinical care and medical advancements.<br>You assess the design, methodology, results, and potential applications of the trials, offering your expert perspective on the significance and relevance of the findings.<br>3) providing a short commentary accompanying their recommendations or evaluations, explaining why they believe the |

| Learning strategy | Prompt |
|---|---|
| | research or trial is important and discussing its potential impact on current understanding, future research, or clinical practice.<br><br>Your answer will contain three parts separated by newlines. Just give the answer with nothing else:<br>A rating on a scale of 1 to 3, representing "Good," "Very Good," and "Exceptional" quality. Return the score only.<br>A short commentary explaining why they are recommending the article and highlighting the potential impact it may have on current understanding and future research.<br>The commentary may include the following considerations: Why the research is noteworthy. How the findings impact current understanding/clinical practice. Any unanswered questions and further discussion points. Other related work (which can be cited) impacts the analysis and evaluation of the article.<br>One or more classifications that provide a quick overview of the reasons why the article is being recommended. These classifications are selected from review, systematic_review, hypothesis, negative, technical_advance, refutation, new_finding, controversial, confirmation, novel_drug_target, good_for_teaching.<br><br>Now please give your answer for this paper:<br>[The target article's abstract] * |
| F-FS and R-FS | You are an expert who provides valuable insights, recommendations, and evaluations related to published research and clinical trial data.<br><br>Specifically, you are responsible for<br>1) Identifying and recommending articles, studies, and research outputs that you consider significant and impactful in the relevant fields.<br>You assess the quality, relevance, and potential impact of the research and provide your expert opinion on its importance.<br>2) evaluating clinical trial results and providing insights into their implications for clinical care and medical advancements.<br>You assess the design, methodology, results, and potential applications of the trials, offering your expert perspective on the significance and relevance of the findings.<br>3) providing a short commentary accompanying their recommendations or evaluations, explaining why they believe the research or trial is important and discussing its potential impact on current understanding, future research, or clinical practice.<br><br>Your answer will contain three parts separated by newlines. Just give the answer with nothing else:<br>A rating on a scale of 1 to 3, representing "Good," "Very Good," and "Exceptional" quality. Return the score only. |

| Learning strategy | Prompt |
|---|---|
| | A short commentary explaining why they are recommending the article and highlighting the potential impact it may have on current understanding and future research. |
| | The commentary may include the following considerations: Why the research is noteworthy. How the findings impact current understanding/clinical practice. Any unanswered questions and further discussion points. Other related work (which can be cited) that impacts the analysis and evaluation of the article. |
| | One or more classifications that provide a quick overview of the reasons why the article is being recommended. These classifications are selected from review, systematic_review, hypothesis, negative, technical_advance, refutation, new_finding, controversial, confirmation, novel_drug_target, good_for_teaching. |
| | Here are six examples with expert recommendations already given, please refer to them when you perform the tasks. |
| | [Abstract 1]<br>1<br>[H1 Connect expert comment for Abstract 1]<br>[Merit codes for Abstract 1] |
| | [Abstract 2]<br>1<br>[H1 Connect expert comment for Abstract 2]<br>[Merit codes for Abstract 2] |
| | [Abstract 3]<br>2<br>[H1 Connect expert comment for Abstract 3]<br>[Merit codes for Abstract 3] |
| | [Abstract 4]<br>2<br>[H1 Connect expert comment for Abstract 4]<br>[Merit codes for Abstract 4] |
| | [Abstract 5]<br>3<br>[H1 Connect expert comment for Abstract 5]<br>[Merit codes for Abstract 5] |
| | [Abstract 6]<br>3<br>[H1 Connect expert comment for Abstract 6]<br>[Merit codes for Abstract 6] |
| | Now please give your answer for this paper: |

| Learning strategy | Prompt |
|---|---|
| | [The target article's abstract] |
| RAG-FS | You are an expert who provides valuable insights, recommendations, and evaluations related to published research and clinical trial data.<br><br>Specifically, you are responsible for<br>1) Identifying and recommending articles, studies, and research outputs that you consider significant and impactful in the relevant fields.<br>You assess the quality, relevance, and potential impact of the research and provide your expert opinion on its importance.<br>2) evaluating clinical trial results and providing insights into their implications for clinical care and medical advancements.<br>You assess the design, methodology, results, and potential applications of the trials, offering your expert perspective on the significance and relevance of the findings.<br>3) providing a short commentary accompanying their recommendations or evaluations, explaining why they believe the research or trial is important and discussing its potential impact on current understanding, future research, or clinical practice.<br><br>Your answer will contain three parts separated by newlines. Just give the answer with nothing else:<br>A rating on a scale of 1 to 3, representing "Good," "Very Good," and "Exceptional" quality. Return the score only.<br>A short commentary explaining why they are recommending the article and highlighting the potential impact it may have on current understanding and future research.<br>The commentary may include the following considerations: Why the research is noteworthy. How the findings impact current understanding/clinical practice. Any unanswered questions and further discussion points. Other related work (which can be cited) that impacts the analysis and evaluation of the article.<br>One or more classifications that provide a quick overview of the reasons why the article is being recommended. These classifications are selected from review, systematic_review, hypothesis, negative, technical_advance, refutation, new_finding, controversial, confirmation, novel_drug_target, good_for_teaching.<br><br>Here are the six semantically similar (6-nearest neighbor) examples with their expert recommendations already given, their 6-NN label for the target paper is 1. Please refer to them carefully when you perform the tasks.<br><br>[Abstract 1]<br>[1/2/3]<br>[H1 Connect expert comment for Abstract 1]<br>[Merit codes for Abstract 1] |

| Learning strategy | Prompt |
|---|---|
| | [Abstract 2]<br>[1/2/3]<br>[H1 Connect expert comment for Abstract 2]<br>[Merit codes for Abstract 2]<br><br>[Abstract 3]<br>[1/2/3]<br>[H1 Connect expert comment for Abstract 3]<br>[Merit codes for Abstract 3]<br><br>[Abstract 4]<br>[1/2/3]<br>[H1 Connect expert comment for Abstract 4]<br>[Merit codes for Abstract 4]<br><br>[Abstract 5]<br>[1/2/3]<br>[H1 Connect expert comment for Abstract 5]<br>[Merit codes for Abstract 5]<br><br>[Abstract 6]<br>[1/2/3]<br>[H1 Connect expert comment for Abstract 6]<br>[Merit codes for Abstract 6]<br><br>Now please give your answer for this paper:<br>[The target article's abstract] |

\* Content in [] are template content that varies for each target article.

In addition to directly examining the correlations between NCS/CSS and model outputs, we further explored an alternative encoding scheme in which CSS is mapped onto the scale of model recommendation outputs, to assess whether this affects the observed correlation patterns. For Task 1, CSS was encoded such that CSS = 0 corresponds to "not recommended" and CSS > 0 corresponds to "recommended." For Task 2, we followed the same mapping as in Figure 11, where CSS levels 1, 2, and 3 correspond to H1 expert ratings of 1, 2, and 3, respectively. The results are reported in Supplementary Table 3.

**Supplementary Table 3. The correlation between model outputs and encoded CSS**

| Task 1[1] (N=1,814) | | Task 2[2] (N=2,572) | |
|---|---|---|---|
| Method | Phi[3] | Method | Spearman[4] |
| Gemini 2.5 Flash Lite (ZS) | 0.1353 | Llama 3.3-70B (R-FS) | 0.0437 |
| GPT-o3-mini (ZS) | 0.2296 | Llama 3.3-70B (ZS) | 0.0476 |

|  | Task 1[1] (N=1,814) |  | Task 2[2] (N=2,572) |  |
| --- | --- | --- | --- | --- |
|  | **Method** | **Phi[3]** | **Method** | **Spearman[4]** |
|  | Llama 3.3-70B (ZS) | 0.2298 | Llama 3.3-70B (F-FS) | 0.0564 |
|  | Qwen3-235B (ZS) | 0.2313 | GPT-4o-mini (F-FS) | 0.0632 |
|  | GPT-o3-mini (RAG-FS) | 0.2316 | Qwen3-235B (ZS) | 0.0685 |
|  | DeepSeek R1 (RAG-FS) | 0.25 | Llama 3.3-70B (RAG-FS) | 0.0726 |
|  | Gemini 2.5 Flash Lite (RAG-FS) | 0.2585 | DeepSeek V3.1 (ZS) | 0.0779 |
|  | Llama 3.3-70B (RAG-FS) | 0.2676 | DeepSeek V3.1 (RAG-FS) | 0.096 |
|  | DeepSeek V3.1 (RAG-FS) | 0.2687 | Gemini 2.5 Flash Lite (R-FS) | 0.1211 |
|  | GPT-5 (RAG-FS) | 0.277 | Gemini 2.5 Flash Lite (F-FS) | 0.1213 |
|  | Llama 3.3-70B (R-FS) | 0.278 | DeepSeek R1 (ZS) | 0.1214 |
|  | Gemini 2.5 Flash Lite (R-FS) | 0.2799 | GPT-4o-mini (ZS) | 0.1248 |
|  | GPT-o3-mini (R-FS) | 0.2823 | GPT-4o-mini (R-FS) | 0.1294 |
|  | GPT-4o-mini (RAG-FS) | 0.2841 | GPT-o3-mini (ZS) | 0.1367 |
|  | GPT-o3-mini (F-FS) | 0.2849 | Gemini 2.5 Flash Lite (RAG-FS) | 0.1378 |
|  | DeepSeek R1 (F-FS) | 0.2885 | Gemini 2.5 Flash Lite (ZS) | 0.145 |
|  | DeepSeek R1 (ZS) | 0.2918 | DeepSeek V3.1 (F-FS) | 0.1463 |
|  | Gemini 2.5 Flash Lite (F-FS) | 0.3107 | DeepSeek R1 (RAG-FS) | 0.1543 |
|  | DeepSeek V3.1 (ZS) | 0.3174 | DeepSeek R1 (R-FS) | 0.1562 |
|  | Qwen3-235B (RAG-FS) | 0.3196 | Qwen3-235B (R-FS) | 0.1591 |
|  | GPT-4o-mini (R-FS) | 0.3203 | GPT-o3-mini (RAG-FS) | 0.1596 |
|  | GPT-5 (ZS) | 0.3265 | Qwen3-235B (RAG-FS) | 0.1606 |
|  | GPT-4o-mini (F-FS) | 0.3316 | GPT-4o-mini (RAG-FS) | 0.1609 |
|  | Llama 3.3-70B (F-FS) | 0.3426 | GPT-o3-mini (R-FS) | 0.1713 |
|  | BERT (SFT) | 0.345 | DeepSeek V3.1 (R-FS) | 0.1955 |
|  | DeepSeek V3.1 (R-FS) | 0.347 | DeepSeek R1 (F-FS) | 0.2059 |
|  | DeepSeek R1 (R-FS) | 0.3492 | Expert (H1 Connect expert rating) | 0.2129 |
|  | RoBERTa (SFT) | 0.3497 | SciBERT (SFT) | 0.2255 |
|  | GPT-4o-mini (ZS) | 0.3614 | Qwen3-235B (F-FS) | 0.2331 |
|  | DeepSeek V3.1 (F-FS) | 0.3712 | GPT-o3-mini (F-FS) | 0.2344 |
|  | GPT-4o-mini (SFT) | 0.3714 | BERT (SFT) | 0.2358 |
|  | Qwen3-235B (F-FS) | 0.3736 | SBERT (SFT) | 0.243 |
|  | BioBERT (SFT) | 0.3778 | PubMedBERT (SFT) | 0.2436 |
|  | SBERT (SFT) | 0.3822 | RoBERTa (SFT) | 0.2453 |
|  | GPT-5 (F-FS) | 0.3829 | RoBERTa-large (SFT) | 0.2501 |
|  | Qwen3-235B (R-FS) | 0.3847 | BioBERT (SFT) | 0.2565 |
|  | RoBERTa-large (SFT) | 0.3848 | GPT-5 (RAG-FS) | 0.2761 |
|  | PubMedBERT (SFT) | 0.385 | GPT-4o-mini (SFT) | 0.3086 |
|  | SciBERT (SFT) | 0.3901 | GPT-5 (ZS) | 0.3177 |
|  | GPT-5 (R-FS) | 0.3965 | GPT-5 (R-FS) | 0.3909 |
|  | Expert (H1 Connect expert rating) | 0.4219 | GPT-5 (F-FS) | 0.4432 |

[1] CSS = 0 – Not recommended, CSS = 1, 2, 3 – recommended.

[2] CSS = 0 removed, CSS = 1 – Good (1), CSS = 2 – Very good (2), CSS = 3 – Exceptional (3)

[3] The Phi coefficient of two binary variables was calculated in this task.

[4] The Spearman coefficient was calculated for two ordinal variables in this task.

Comparing Task 1 correlation results with those in Figure 5, we can still observe largely consistent patterns: the expert label leads both analyses, fine-tuned models (GPT-4o-mini SFT, BERT family) outperform prompt-based LLMs in both, and GPT-o3-mini is consistently weakest. However, binarizing CSS into 0 and 1, 2, 3 substantially compresses the spread of correlations. Further, for Task 2 compared against Figure 9, restricting to CSS > 0 barely changes method rankings or values. Expert correlation remains low (~0.22) in both analyses. GPT-5 dominates clearly (Spearman up to 0.47), fine-tuned BERTs form a second tier (~0.22–0.26), and most prompt-based LLMs are weak (0.04–0.20).